\documentclass[preprint,12pt,1p,a4paper]{elsarticle}
\usepackage{amsmath}
\usepackage{amssymb}
\usepackage{longtable}
\usepackage{pifont} 
\usepackage{csp} % my version!
\usepackage{varwidth}
\usepackage{mathtools,bm}
\newlength\mylen
\usepackage{array}
\usepackage{svg}
\usepackage{adjustbox}

\makeatletter
\DeclareOldFontCommand{\rm}{\normalfont\rmfamily}{\mathrm}
\DeclareOldFontCommand{\sf}{\normalfont\sffamily}{\mathsf}
\DeclareOldFontCommand{\tt}{\normalfont\ttfamily}{\mathtt}
\DeclareOldFontCommand{\bf}{\normalfont\bfseries}{\mathbf}
\DeclareOldFontCommand{\it}{\normalfont\itshape}{\mathit}
\DeclareOldFontCommand{\sl}{\normalfont\slshape}{\@nomath\sl}
\DeclareOldFontCommand{\sc}{\normalfont\scshape}{\@nomath\sc}
\makeatother

\newcommand{\xmark}{\ding{55}}%                                                                                  
\newcommand{\cspm}{CSP$_{\textrm M}$}
\newcommand{\ind}[1]{\hspace*{#1.0\parindent}}

\title{Verifying Correctness of Shared Channels in a Cooperatively Scheduled Process-Oriented Language}

\author[1]{Jan Pedersen\corref{cor1}}
\ead{matt.pedersen@unlv.edu}

\author[2]{Kevin Chalmers}
\ead{k.chalmers@rave.ac.uk}

\cortext[cor1]{Corresponding Author}

\affiliation[1]{organization={Department of Computer Science, University of Nevada Las Vegas},
addressline={4505 South Maryland Parkway},
postcode={89154},
city={Las Vegas, NV},
country={USA}}

\affiliation[2]{organization={School of Computing, Architecture, and Emerging Technologies, Ravensbourne University},
addressline={6 Penrose Way},
postcode={SE10 0EW},
city={London},
country={UK}}

\begin{document}

\begin{abstract}
Correct concurrent behaviour is important in understanding how components will act within certain conditions. In this work. we analyse the behaviour of shared communicating channels within a coorporatively scheduled runtime. We use the refinement checking and modelling tool FDR to develop both specifications of how such shared channels should behave and models of the implementations of these channels in the cooperatively scheduled language ProcessJ. Our results demonstrate that although we can certainly implement the correct behaviour of such channels, the outcome is dependant on having adequate resources available to execute all processes involved. We conclude that modelling the runtime environment of concurrent components is necessary to ensure components behave as specified in the real world.
\end{abstract}

\begin{keyword}
CSP, process-orientation, processJ, formal verification, cooperatively scheduler, verified runtime
\end{keyword}

\maketitle
\newpage
%!TeX root = ./Shared.tex
\section{Introduction}
\label{sec:introduction}

The correctness of software is an important issue. There are two main approaches to ensuring correctness: testing and formal verification. Almost all software is tested (more or less) rigorously until such time that the developer is convinced that most of the serious bugs have been eradicated. However, testing is not always enough. In~\cite{PedersenWelch18} a simple piece of software for a made-up Martian rover was considered. The software could end up in a deadlocked configuration (because of an inversion error) but the authors calculated that ``...we would need to wait towards 8 million seconds (over 90 days) to reach a 50\% chance of seeing that deadlock triggered.'' This illustrates that using testing alone to ensure the correctness of software is simply not enough. The Mars rover system from above was formally verified using CSP~\cite{Ho78, Ho85} (Communicating Sequential Processes) and the formal model checker FDR~\cite{fdr}, and the inversion error was found immediately and FDR produced an execution trace explaining how the error could occur. This is just one example of how formal verification can find errors and bugs that testing alone may never encounter.  Formally verified correct software becomes even more important when we consider runtime systems associated with programming languages: since a large number of programs will be built upon the language's runtime, it is vitally important that it (the runtime) is bug free and behaves according to its specification. 

In this paper we utilise CSP, a process algebra defined using processes, channels, and choice. It supports process composition using a parallel and interleaving operators. Section~\ref{sec:CSP} goes into more detail about CSP and the various semantic models that can be used to check refinement between processes. This paper models the implementation of shared channels in the ProcessJ runtime, demonstrating their correctness against a well-formed and proved specification. However, we also present limitations encountered due to the limitations of scheduling.

ProcessJ is a process-oriented programming language being developed at the University of Nevada Las Vegas and the University of Roehampton. ProcessJ has non-buffered synchronous channels for communication and is highly concurrent. It is cooperatively scheduled, running on the JVM. This means that the runtime system for ProcessJ is written in Java; that is, the code that handles scheduling and communication etc. is implemented in Java. It is extremely important that this code (we will collectively refer to as ``the runtime'') is correct---that is, it should never deadlock or livelock and it should behave according to a predetermined specification describing what acceptable behaviour is. In~\cite{FormaliSE19} we started out this verification journey to prove the behaviour for the ProcessJ runtime correct. In this initial approach we translated the ProcessJ runtime code for channel communication to CSP and proved it to behave correctly when comparing to a well-established specification. In~\cite{PedersenChalmers} we added verification of choice and realised that the {\it standard approach} used for verification was insufficient---it fails to take the code for the runtime system (i.e., schedulers, runners etc.) into account. We showed that this is crucial to do as the behaviour of a {\it scheduled runtime} becomes limited in terms of possible behaviours. With this in mind we redid the verification of the existing scheduler of ProcessJ and realised that there were some issues with the implementation (live-lock in the scheduler). These issues were tackled in~\cite{ChalmersPedersen24} and we showed that both one-to-one channel communication and alternation (choice) now work correctly with the the new scheduling system.

Now, with a well-behaved scheduling and execution system, it is time to attack the verification of the remaining parts of the runtime system. In this paper we consider shared channels; that is, channels with multiple readers and/or multiple writers. 

In~\cite{FormaliSE19} and~\cite{PedersenChalmers} we started with an actual implementation which we translated to CSP in order to perform the verification. In~\cite{ChalmersPedersen24} we started with a specification and then implemented the verified scheduler in Java. In this paper we again switch back to translating existing code to CSP for verification (using the new verified scheduling system).

\subsection{Motivation}

A shared channel is one where multiple readers and/or writers may be engaged with a channel. The individual readers/writers do not synchronise amongst themselves and only synchronise with one partner (i.e., one reader and one writer synchronise) during communication. Such behaviour is allowed in CSP and an example is provided in Section~\ref{sec:spec}. Due to implementation challenges when considering choice, the approach is to provide shared channels in CSP inspired implementations that differ from one-to-one channels. In ProcessJ, we do not rely on preemptive scheduling and thus have implemented a queuing mechanism for mutual exclusion on the shared channel ends. Our aim is to demonstrate that our implementation behaves as would be expected for such a channel.

In the standard approach to channel-based system verification, no runtime system, execution system, hardware etc. is modelled. Events in the formal model can happen whenever the system is ready to perform the event. The verification is in the abstract rather than in a concrete environment. We argue that such an approach does not model what happens when a channel-based system is run in the real-world. The standard way of modelling such systems assumes that a ready event will remain available until scheduled. However, this also assumes the extreme case---that an event may immediately occur, implying resources are available to execute the event when it becomes available. Consider a process that is at the end of a queue of ready processes. It must wait to be scheduled---its events will not be immediately available. This assumption that resources are not scarce is not true. By modelling the execution environment as we do in this and previous papers, we are much closer to the behaviour of a real running implementation of a channel.

\subsection{Contribution}

The main contributions of this paper fall into three parts:
\begin{itemize}
\item We have verified that the ProcessJ implementation of shared channels using the existing cooperating scheduler in the ProcessJ runtime is correct. That is, they behaves as expected by virtue of passing refinement checks using a CSP translation of the implementation and a CSP generic model of a shared channel.
\item We have developed an extended algebraic proof that the shared channel specification indeed does behave like a CSP shared channel.
\item Finally, we have further demonstrated the relationship between available resources and active processes for meeting specifications. Again, we have shown that in order to obtain refinement between specification and model in both directions, the number of available schedulers\footnote{Note, a {\it scheduler} is really a {\it runner}; something that executes a ProcessJ process. We have in prior papers referred to these entities as schedulers, and carry on with that nomenclature in this paper as well.} must be equal to or larger than the number of processes in the system.\footnote{Or at least enough in the way that any ready process can be run immediately.}
\end{itemize}

\subsection{Channel Communication}

In a process-oriented language like ProcessJ, communication between processes happens though non-buffered synchronous channels. The sending process performs a blocking write with a value on the channel by using the {\it writing end}. The reading process performs 
a blocking read on the {\it reading end}. When both the writer and the reader are ready, they synchronise and the value is exchanged. In~\cite{PedersenChalmers,ChalmersPedersen24} we considered the correctness of one-to-one channels; that is, channels with exactly one sender and exactly one receiver. In this paper we introduce channels with shared channel ends to ProcessJ. This includes channels with a single writing end and multiple reading ends; multiple writing ends and a single reading end; and multiple writing ends and multiple reading ends. 

A one-to-one channel carrying values of type $type$ (named $c$) in ProcessJ is declared as follows:\\

\noindent
\ind1{\bf chan}$<\!type\!>\ c$;\\

\noindent
Channels with shared ends are declared similarly but the keyword {\bf shared} is used to denote the end which is shared:\\

\noindent
\ind1{\bf shared write chan}$<\!type\!>\ sharedWriteChan$;\\
\ind1{\bf shared read chan}$<\!type\!>\ sharedReadChan$;\\
\ind1{\bf shared chan}$<\!type\!>\ sharedWriteAndReadChan$;\\

\noindent
reading and writing to a non-shared or shared channel end is similar; here are both reading and writing:\\

\noindent
\ind1$c.${\bf read}()\\
\ind1$c.${\bf write}$(expr)$\\

When using a shared end of a channel, the ProcessJ compiler generates the correct protection to avoid concurrent access and other race conditions. In this paper we validate this compiler-generated runtime code along with the runtime scaffolding like access to queues of reader and writers all waiting to use a shared channel end.
A more comprehensive example of a one-to-many (shared reading end) and a many-to-one (shared writing end) can be found in Figure~\ref{fig:sharedEnds}. A simple one-to-one channel use can be found in Section~\ref{sec:ProcessJ}

\subsection{Breakdown of the Paper}

In Section~\ref{sec:background} we present a short discussion about scheduling as well as present related and prior work. Section~\ref{sec:ProcessJ} presents the ProcessJ programming language, and Section~\ref{sec:CSP} gives a brief introduction to Communicating Sequential Processes (CSP). In Section~\ref{sec:SchedulingAndProcesses} we lay the groundwork for this paper by introducing a model for a scheduler and a process.  We then continue on in Section~\ref{sec:sharedChansInPJ} by illustrating the use of and translation in to Java of ProcessJ shared channels. This translation into Java is further translated into CSP in Section~\ref{sec:imp}.
A generic channel supporting multiple readers and writers is specified in CSP in Section~\ref{sec:spec} and a formal proof this specification is given in Section~\ref{sec:proof}.
Finally, Sections~\ref{sec:results} and~\ref{sec:discussion} contain the results and a discussion of them. We close with future work in Section~\ref{sec:futureWork}.

The CSP code presented in this paper, along with all the tested assertions, can be found in the GitHub repository located at {\tt http://github.com/} {\tt mattunlv/process-shared-csp}.    % 1. Introduction    
%!TeX root = ./Shared.tex
% Section 2
\section{Background Related Work}  
\label{sec:background}

In this section we start with some background information about cooperative scheduling and then consider the results from~\cite{PedersenChalmers}. 

\subsection{Cooperative Scheduling and Multitasking}

To support multi-processing, a runtime typically takes one of two scheduling approaches:

\begin{itemize}
	\item {\em preemptive scheduling}---processes are allocated processor time-slices managed centrally (e.g., via the operating system).
	\item {\em cooperative scheduling}---processes decide when to give up ({\em yield}) the processor.
\end{itemize}

With no priorities, preemptive scheduling guarantees processes will be allocated processor time. Cooperative scheduling makes no such guarantee as a process may use resources indefinitely without yielding. In both cases, a method is required to determine the next process to service. Typically, and in the case of ProcessJ, a queue of processes is maintained.

On a single processor system, we can use a simple approach to cooperative scheduling---a continuous dequeing of the next process to execute until the queue is empty. In a multiprocessor system, consideration of cross-core synchronisation is required. Often, and the case in ProcessJ, we rely on preemptive locking mechanisms provided by the operating system to ensure correct concurrent behaviour. We therefore have to account for the cooperative scheduling (via yield points) within the preemptive scheduling (via locking) when designing a system. Thus, for ProcessJ runtime verification, we {\bf must} incorporate the implementation of the cooperative scheduling environment.

There are advantages to cooperative scheduling. For example, in ProcessJ, processes are stackless; a process' local data is stored as object data. Therefore, a process is incredibly lightweight. Previous ProcessJ publications~\cite{ShresthaPedersen16} have demonstrated hundreds of millions of processes in operation. This compares to preemptive thread-based libraries, such as JCSP, supporting only a few thousand processes. Large stack sizes (often greater than a megabyte) that each thread utilises increases memory requirements.

ProcessJ also has fast context switching due to its lightweight processes~\cite{ShresthaPedersen16}. As no preemptive thread swapping is required, context switches, and thus channel communication time, is fast. A thread provided by the operating system will save a register record and load a new one onto the CPU with each context switch. If this switch is cross-core, the impact is greater due to the stack swapping between caches.

Cooperative {\it scheduling} has limited popularity due to the programmer typically having to manage scheduling. Operating systems favour preemptive scheduling. Threading libraries in programming languages typically use operating system provided primitives. There are some exceptions (e.g., the Win32 fiber library), but preemptive scheduling is the {\it de facto} approach to threading. No mainstream language directly supports cooperative scheduling as part of its standard library.

We believe that hiding cooperative scheduling (via communication), as in ProcessJ, will alleviate some of the challenges in managing such scheduling. Indeed, until version 1.10, the Go scheduler was a mix of cooperative and preemptive. The current Go scheduler is preemptive, but Go is a compiled language that runs directly on hardware. As the aim of ProcessJ is to run within the JVM for portability, we are limited in how fast and lightweight processes are supported. {\sf occam}-$\pi$ also featured a cooperative scheduler~\cite{ritson2009} which was likewise hidden from the programmer.

In contrast, cooperative {\it multitasking} has a growing popularity, primarily from the {\it async/await} pattern. Coroutines~\cite{Moura2009} have gained popularity in languages ranging from JavaScript and Python to C++ and Rust. Indeed, this has led to the discussion of new concurrency paradigms such as structured concurrency~\cite{Chen2022} which have influenced thinking in Kotlin and Swift. However, coroutines are not scheduled as in the case of a cooperative runtime as used in ProcessJ.

The ProcessJ scheduler is fairly unique. Other lightweight concurrency languages, such as Go and Erlang, typically use preemptive scheduling~\cite{Chabbi2022, cesarini2009erlang}. Indeed, the only similar runtime approach---hiding cooperation in synchronisation---we have found is in Stackless Python~\cite{tismer2000continuations}.

\subsection{Related Work}
\label{sec:relatedWork}

Verification of runtimes has been undertaken for other systems like CSO~\cite{sufrin2008communicating} and JCSP~\cite{Welch2010} as well as the regular non-shared channels and choice for ProcessJ~\cite{PedersenChalmers}. Similarly, other systems have been formally verified using CSP/FDR or other similar verification systems like SPIN (for example see~\cite{namvaritazehkand2024novel}).

Mutual-exclusion is long-solved (e.g., see~\cite{Lamport87}). Model checking concurrent algorithms often begins with the assumption that multi-processor behaviour is typically managed via a preemptive scheduler. Atomic operations introduce new challenges for mutual exclusion, but again, the preemptive scheduler enables certain assumptions (e.g., see~\cite{michael1995simple}). The assumption of preemptive scheduling has been successfully applied to model check synchronous channel communication~\cite{WelchMartin00, Lowe2019} in JCSP~\cite{Welch2010} and Communicating Scala Objects (CSO)~\cite{sufrin2008communicating}. JCSP and CSO both rely on preemptive scheduling provided via Java threads. The ProcessJ scheduler more closely emulates the {\sf occam}-$\pi$ multi-processor cooperative scheduler defined by Ritson et. al.~\cite{ritson2009}.

However, as we emphasized in~\cite{PedersenChalmers}, all of those previous attempts have overlooked the \textit{execution environment} (scheduler, hardware processors, etc.) when performing the verification. The verification that we performed in~\cite{PedersenChalmers} was done not only in the standard way (not taking into account the execution environment), but also taking into account the cooperative scheduler. This is how we demonstrated the livelock in the original ProcessJ scheduler.

\subsection{Our Work So Far}
In~\cite{FormaliSE19} and~\cite{PedersenChalmers} we performed verification of channel communication using one-to-one channels, first in the standard unrestricted (non-scheduled) manner, and then in a restricted manner, where the code of the runtime system and the scheduler was taken into account. In the unrestricted (non-scheduled version) we achieved the results that we expected. We then introduced a simple scheduler:

\noindent
\begin{tabbing}
\ind1$Queue\!\!<\!\!Process\!\!>\ runQueue$;\\
\ind1$\ldots$\\
\ind1// enqueue 1 or more processes to run ... \\
\ind1{\bf while} (!$runQueue.isEmpty$()) \{ \\    
\ind2$Process\: p = runQueue.dequeue()$; \\  
\ind2{\bf if} ($p.ready()$)\\
\ind3$p.run$();\\
\ind2{\bf if} ($!p.terminated$())\\
\ind3$runQueue.enqueue(p)$;\\
\ind1\}
\end{tabbing}

When verifying our code taking this scheduler into account, the verification software determined that our code could livelock. This was caused through the run queue containing both ready and non-ready processes, and with more than one scheduler, it could be possible that one of more of the schedulers never executed any code but simply dequeued and re-enqueued non-ready processes forever. In~\cite{ChalmersPedersen24} we fixed this issue by implementing an improved scheduler and removing the non-ready processes from the run queue. We could now perform the verification again and get the results we were aiming for. The aim for this paper is to succeed in the same manner with a scheduled system using channels with shared ends.      % 2. Background and Related work
%!TeX root = ./Shared.tex
% Section 3
\section{ProcessJ} 
\label{sec:ProcessJ}

ProcessJ~\cite{ShresthaPedersen16} is a process-oriented programming language being developed at the University of Nevada Las Vegas. The syntax of ProcessJ resembles Java, with new constructs for synchronisation. It has CSP semantics similar to {\sf occam}/{\sf occam}-$\pi$~\cite{occam,occam-pi-home}. The ProcessJ compiler generates Java source files that are further compiled to produce class files, which in turn are rewritten using the ASM~\cite{ASM} bytecode rewriting tool to create cooperatively schedulable Java bytecode. Targeting the JVM ensures maximum portability, and tests have shown that a computer can handle upwards of half a billion processes on a single core~\cite{ShresthaPedersen16}.

ProcessJ supports four different channel types: one-to-one, one-to-many, many-to-one, and many-to-many. In this paper we tackle the correctness of the last three channel types.

Below is a simple ProcessJ example (with a single one-to-one channel for simplicity). The {\it main} method starts two processes ({\it reader} and {\it writer}) concurrently; {\it reader} is passed the reading end of a channel, and {\it writer} is passed the writing end of the same channel. The writing process sends the value 42 to the reading process, which in turn prints out the value. Once both these processes have terminated, the {\it main} process prints ``Done''.

\noindent
\begin{tabbing}
\hspace*{7cm} \= \\ \kill
\hspace*{1\parindent}{\bf public void} {\it reader}({\bf chan}$<${\bf int}$>$.{\bf read} {\it in}) \{ \> \\
\hspace*{2\parindent}{\bf int} $x$ = $in$.{\bf read}(\ ); // read a value from the {\it in} channel\\
\hspace*{2\parindent}{\bf println}(``received:'' + $x$);   // print the received value\\
\hspace*{1\parindent}\} \> \\
$ $ \> \\
\hspace*{1\parindent}{\bf public void} {\it writer}({\bf chan}$<${\bf int}$>$.{\bf write} {\it out}) \{ \> \\
\hspace*{2\parindent}$out$.{\bf write}(42); // write the value 42 on the {\it out} channel\\
\hspace*{1\parindent}\} \> \\
$ $ \> \\
\hspace*{1\parindent}{\bf public void} {\it main}({\bf string} args[\ ]) \{ \> \\
\hspace*{2\parindent}{\bf chan}$<${\bf int}$>$ $c$;  // declare a channel\\
\hspace*{2\parindent}{\bf par} \{ // in parallel, run the following two processes: \\
\hspace*{3\parindent}$writer$($c$.{\bf write}); // a writer w/ the writing end\\
\hspace*{3\parindent}$reader$($c$.{\bf read}); // a reader w/ the reading end\\
\hspace*{2\parindent}\} \> \\
\hspace*{2\parindent}{\bf println}(``Done''); \> \\
\hspace*{1\parindent}\} \> \\
\end{tabbing}

\subsection{Processes}
\label{sec:processes}

The ProcessJ compiler outputs Java source code that links with a runtime systems. Every ProcessJ procedure becomes a subclass of the following ProcessJ runtime {\it PJProcess} class:\\

\noindent   
\hspace*{1\parindent}\textbf{class} \textit{PJProcess} \{ \\
\hspace*{2\parindent}\textbf{int} \textit{runLabel} = 0;\\
\hspace*{2\parindent}\textbf{bool} \textit{ready} = true;\\
\hspace*{2\parindent}\textbf{bool} \textit{running} = false;\\
$ $\\
\hspace*{2\parindent}\textbf{public} \textbf{synchronized} \textbf{void} $setNotReady$() \{\\
\hspace*{3\parindent}$ready$ = {\bf false}; // Set the process not-ready\\
\hspace*{2\parindent}\}\\
\hspace*{2\parindent}\\ 
\hspace*{2\parindent}{\bf public synchronized void} $setReady$() \{\\
\hspace*{3\parindent}$ready$ = {\bf true}; // Set the process ready\\
\hspace*{2\parindent}\}\\
\hspace*{2\parindent}\\
\hspace*{2\parindent}\textbf{public void} \textit{run}() \{ \}\\  
$ $\\
\hspace*{2\parindent}// Dummy methods for ASM\\
\hspace*{2\parindent}\textbf{public void} $yield$(\textbf{int} $label$) \{ \}\\
\hspace*{2\parindent}\textbf{public void} $label$(\textbf{int} $label$) \{ \}\\
\hspace*{2\parindent}\textbf{public void} $resume$(\textbf{int} $label$) \{ \} \\
\hspace*{1\parindent}\}\\

\noindent
\noindent
$setNotReady()$ and $setReady()$ set a process {\em not-ready-to-run} and {\em ready-to-run}, respectively.  For example, the $reader$ procedure would generated the following code:

\begin{tabbing}
\ind1{\bf public class} $reader$ {\bf extends} $PJProcess$ \{\\
\ind2{\bf private int} $x$; // local ProcessJ variable {\it x}\\
$ $\\  
\ind2{\bf public void} $run$() \{\\
\ind3  {\bf switch}($runLabel$) \{\\
\ind3  {\bf case} 0: $resume$(0); {\bf break};\\
\ind3  ...\\
\ind3  \}\\
$ $\\
\ind3// code for $x$ = $in$.{\bf read}();\\
\ind3// code for {\bf println}({\tt "received:"} + $x$);\\
\ind2\}\\
\ind1\}
\end{tabbing}

\noindent
All local variables of $reader$ are transformed into fields of the class; this way there is no need to maintain a stack of locals between invocations of the $run()$ method. The process-flow of the ProcessJ compiler is as follows:

\begin{itemize}
\item The ProcessJ compiler reads a ProcessJ source file and produces, as output, a number of Java source files; one per ProcessJ procedure. This code contains invocations of three dummy methods: $yield()$, $resume()$, and $label()$, that act as placeholders for the later instrumentation phase.
\item The generated Java source files are compiled with the Java compiler and class files are produced.
\item The three dummy methods $yield()$, $resume()$, and $label()$ are used and transformed in the generated code as follows:
\begin{itemize}
\item $resume(i)$ is replaced by a `{\tt goto} address of $label(i)$' when the bytecode rewriting is performed.
\item $label(i)$ is used to obtain the actual address of that particular location in the code. During bytecode rewriting it is replaced by a {\tt nop} instruction once the address has been recorded.
\item $yield(i)$ is replaced by a jump to the end of the code after setting the $runLabel$ to $i$. This way, when $run()$ is called again, the switch at the start of the code will jump to the label associated with the current $runLabel$. 
\end{itemize}
\item The resulting instrumented bytecode is now compatible with the ProcessJ cooperative scheduler and can now be executed.
\end{itemize}

Each process has the following code at the beginning of its {\it run()} method:\\

\noindent
\hspace*{1\parindent}{\bf switch} ({\it runLabel}) \{\\
\hspace*{2\parindent}{\bf case} 1: {\it resume}(1); {\bf break};\\
\hspace*{2\parindent}{\bf case} 2: {\it resume}(2); {\bf break};\\
\hspace*{2\parindent}$\ldots$\\
\hspace*{2\parindent}{\bf default}: // runtime error.\\
\hspace*{1\parindent}\}\\

\indent
Along with setting the $runLabel$ for $yield(i)$, this code implements the core of the cooperative scheduling from a process' point of view, namely, the ability to jump back to where the process last left off (called a $yield()$).

%The full implementation of the ProcessJ runtime can be found at the ProcessJ GitHub repository at {\tt https://github.com/mattunlv/ProcessJ} and the CSP developed in this paper is also available through GitHub at {\tt https://github.com/mattunlv/processj-csp

%As an example operation, consider a channel communication for which the writer arrives first. As the reader is not present, the writer can copy its data to the channel, register itself as the writer on the channel, and wait for the reader to appear. Waiting is performed by yielding, which sets a {\it run label} used for resumption when the process is rescheduled.  The writer also sets itself not-ready to run before returning to the scheduler. While the writer is not ready, the scheduler will not resume it---it simply gets put at the back of the run queue again. Only when a reader appears is the writer set to ready. At this time, the scheduler is free to resume the writer when it reaches the front of the run queue again. 

In Section~\ref{sec:sharedChansInPJ} we look more closely at the Java code that gets generated for both channel reads and writes.        % 3. ProcessJ
%!TeX root = ./Shared.tex

\section{Communicating Sequential Processes}
\label{sec:CSP}

Communicating Sequential Processes (CSP)~\cite{Ho78,Ho85,theoryAndPractice,awr135} is a process algebra enabling concurrent system specification via {\em processes} and {\em events}. Processes are abstract components defined by the events they perform.

Events are {\em atomic}, {\em synchronous}, and {\em instantaneous}. That is, events are indivisible, and cause all engaged processes to wait until the event occurs, and when the event occurs it is immediate to all engaged processes. A CSP model defines the events processes are willing to synchronise upon at different stages of their execution.

The simplest CSP processes are {\sf STOP}---which engages in no events and will not terminate---and {\sf SKIP}---which engages in no events but will terminate.

Events are added to a process using the {\em prefix} operator ($\then$).  For example, the process $P = x \then$ {\sf STOP} engages in event $x$, then stops.  A process definition must end by executing another process definition; that is the general form is {\it Process} = {\it event} $\then$ {\it Process'}. Processes can be recursive (e.g., $P = x \then P$), with an anonymous form (e.g., $\mu r.x \then r$) we will use for simplicity in the proof of our shared channels in Section~\ref{sec:proof}.

\subsection{Choice}

CSP has several {\em choice} operators to exhibit branching behaviour. The three most frequently used choice types are {\em external (or deterministic) choice}, {\em internal (or non-deterministic) choice}, and {\em prefix choice}.

Given two processes $P$ and $Q$, the definition $P \extchoice Q$ (external choice) is a process that will behave as $P$ or $Q$ based on the first event offered by the environment. For example, the process:

\begin{tabbing}
=\===\===\=\kill	
\>$P = (a \then Q) \extchoice (b \then R)$
\end{tabbing}

\noindent
is willing to accept $a$, then behave as $Q$, or accept $b$, then behave as $R$.  The system can (non-deterministically) choose either when both $a$ and $b$ are available.

An internal choice is represented by $\intchoice$. A process $P \intchoice Q$ can behave as either $P$ or $Q$ without considering the external environment. 

Both external choice and internal choice can operate across a set of events. If $E=\{e_1,\ldots,e_n\}$ is a set of events, then 

\begin{tabbing}
=\===\===\=\kill	
\>$\underset{a \in E}{\Extchoice} a \then P \equiv e_1 \then P\extchoice e_2 \then P \extchoice \cdots \extchoice e_n \then P$ \\
\>$\underset{a \in E}{\Intchoice} a \then P \equiv e_1 \then P \intchoice e_2 \then P\intchoice \cdots \intchoice e_n \then P$
\end{tabbing}

With prefix choice we can define an event and a parameter. If we define a set of events as $\{ c.v \ | \ v \in Values\}$, we can consider $c$ as a channel willing to communicate a value $v$.  We can then define input and output operations ($?$ and $!$, respectively) to allow communication and binding of variables. This is simply a shorthand due to the following identities:

\begin{tabbing}
=\===\===\=\kill	
\>$c!v \then P \equiv c.v \then P$ \\
\>$c?x \then P \equiv \underset{\mathclap{x \ \in Values}}{\Extchoice} \ c.x \then P$
\end{tabbing}

\noindent
where {\it Values} is a finite set. 

CSP supports a functional {\it if} statement with each branch ending in a process definition. It is therefore common to write:

\begin{tabbing}
=\===\===\=\kill	
\>1${\sf if}\ (b)\ {\sf then} \ P$\\
\>${\sf else} \ SKIP$
\end{tabbing}

\noindent
The semantics of $c?v \then ({\sf if}\ (v == x)\ {\sf then}\ P\ {\sf else}\ Q)$ in CSP is equivalent to $(c.x \then P) \extchoice (\underset{\mathclap{v \in X - \{ x \} }}{\Extchoice} \ c.v \then Q)$\footnote{Assuming that $v$ does not appear in $P$.}. 

\subsubsection{Pre-Guards}

It is possible to control the availability of choice branches (i.e., whether the branch will be considered) by placing a Boolean expression and a \& before the choice branch as a pre-guard. For example:

\begin{tabbing}
=\===\===\=\kill	
\>$e_1\ \&\ c?x \then \ldots\ \extchoice\ e_2\ \&\ d?x \then \ldots$ 
\end{tabbing}

\noindent
If the Boolean expression $e_1$ is true and the Boolean expressions $e_2$ is false only the $c?x$ guard will be considered. Similar, if only $e_2$ is true, only $d?x$ will be considered; only if both $e_1$ and $e_2$ are true will both guards be considered in the alternation. 

\subsection{Process Composition}

Processes can be combined via {\em parallel} and {\em sequential} composition. A parallel composition is denoted as $P \ || \ Q$. $P$ and $Q$ must now synchronise on a shared set of events. There are two forms of parallel composition.

\begin{itemize}
	\item The generalised parallel $P \  \underset{A}{||} \ Q$ defines two processes with a {\it synchronisation set} $A$.
	\item The alphabetised parallel $P \ _{A} || _{B} \ Q$ defines two processes that can only perform events in their respective {\it alphabets}; $P$ with alphabet $A$ and $Q$ with alphabet $B$.
\end{itemize}

For the generalised parallel, both $P$ and $Q$ must offer any event in $A$ at the same time for that event to occur. Events not in $A$ will not cause synchronise between $P$ and $Q$, although $P$ and $Q$ may still perform such events. For example,

\begin{tabbing}
=\===\===\=\kill	
\>$(a \then b \then P) \ \underset{ \{ b \} }{||} \ (b \then c \then Q)$
\end{tabbing}

\noindent
has two processes synchronising on $b$. Thus, $a$ must happen, before both processes execute $b$ and then $c$ is performed. If the synchronisation set also contained $c$, then the right-hand side would not be able to perform this event as the left-hand side would not offer it.

For the alphabetised parallel, $P$ and $Q$ must synchronise on any events in $A \cap B$. For example,

\begin{tabbing}
=\===\===\=\kill	
\>$(a \then b \then P) \ _{ \{ a, b \} } || _{ \{ b, c \} } \ (b \then c \then Q)$
\end{tabbing}

\noindent
has two processes synchronising on the intersection alphabet $\{ b \}$. Thus, $a$ must happen, before both processes execute $b$ and then $c$ is performed. If the alphabet on the left-hand side did not contain $a$, then system would deadlock as $a$ could not be performed.

Two processes can also {\em interleave}: $P \ ||| \ Q$, which means that $P$ and $Q$ execute in parallel but {\em do not} synchronise on any shared events. Returning to our previous example, $(a \then b \then P) \ ||| \ (b \then c \then Q)$ may accept $a$ or $b$ first. The event $b$ is only ever performed by one process at a time. As with choice, we can interleave over a set, such as $\underset{\mathclap{a \in \{ 1 \dots n\} }}{|||} \ P$ to run $n$ interleaving instances of process $P$.

Sequential composition is denoted as $P\ ; Q$. This means after $P$ has terminated, $Q$ is performed. Therefore, it allows the definition of behaviour as a sequential composition of discrete process definitions.

\subsection{Traces and Hiding}

The {\it trace set} is the set of all a process's externally observed sequences of events . For example, for a process $P = a \then P$ the shortest trace observable is the empty trace $\trace{}$. When $P$ engages on the event $a$ for the first time we can observe this external event and we now have a trace $\trace{a}$. The second time $P$ engages on $a$ we can observe the trace $\trace{a,a}$, and we say that the traces of $P$ are $\Traces(P) = \{\trace{},\trace{a},\trace{a,a},\ldots\}$.  For a process $Q = a \then b \then Q$ has a trace-set $\{ \trace{}, \trace{a}, \trace{a,b}, \trace{a,b,a}, \dots \}$. 

Observable process events can be concealed via the hiding operator $\backslash$. Hidden events are replaced by $\tau$ and are ignored when comparing observable traces. For example, $(a \then b \then a \then {\sf SKIP}) \backslash \{a\}$ has traces $\{\trace{}, \trace{b} \}$ as $\tau$s are not observable. Similarly, $\Traces(P\hide\{a\}) = \{\trace{}\}$ and $\Traces(Q\hide\{a\}) = \{ \trace{}, \trace{b}, \trace{b,b}, \ldots\}$.

\subsection{Models}
\label{sec:models}

CSP defines three semantic models of behaviour to analyse systems: the {\it traces} model, the {\it failures} model, and the {\it failures/divergence} model. Each model extendeds analysis upon the previous model to establish stronger refinement checks.

\subsubsection{The Traces Model}

The {\bf traces model} comes from the externally observed behaviour of a system. If the trace set of a system {\em implementation} $Q$ is a subset of the trace set of a system {\em specification} $P$ (i.e., $\Traces(Q) \subseteq \Traces(P))$, we state that $P \refinedby[T] Q$ ($Q$ {\em trace refines} $P$). In a refinement test {\it Specification} $\refinedby[T]$ {\it Implementation}, {\it Specification} can be thought of as the allowable traces. The refinement test checks if an {\it Implementation} has only allowable traces.

Hiding events allows us to analyse the external behaviour of a process, thus allowing assertions such as $P \refinedby[T] Q \land Q \refinedby[T] P$. An implementation may contain events that are not part of the specification. These are events required to model the implementation of the system rather than just a specification of behaviour. We therefore hide internal events of an implementation to enable refinement checking.

\subsubsection{The Stable Failures Model}

The {\bf stable failures model}~\cite{theoryAndPractice,awr135} deals with events that a process may refuse to engage in after a trace. A stable state is one where a process $P$ cannot make internal progress (i.e., via hidden events) and must engage externally. A refusal is an event a process cannot engage with when in a stable state.

The stable failures model overcomes the limitation of comparing traces.  For example, $(P = a \then P) \refinedby[T] ((P = a \then P) \intchoice {\sf STOP}))$, although the right-hand definition may non-deterministically refuse to accept any events. A {\em failure} is a pair $(s, X)$ where $s$ is a trace of a process $P$, and $X$ is the set of events that can be refused after $P$ has performed the trace $s$.  Stating that $P \refinedby[F] Q$ means that whenever $Q$ refuses to perform a set of events, $P$ does likewise. More formally, $P \refinedby[F] Q \Leftrightarrow failures(Q) \subseteq failures(P)$. Therefore, a specification in the stable failures model defines which failures an implementation is allowed to have.

\subsubsection{The Failures-Divergences Model}

The final model is the {\bf failures-divergences model}.  Divergences deal with potential livelock scenarios where a process can continually perform internal events and not progress in its externally observed behaviour. For example, consider the following two processes:

\begin{tabbing}
	\noindent
	\ind1$P = a \then {\sf STOP}$ \\
	\ind1$Q = (a \then {\sf STOP}) \intchoice {\sf DIV}$
\end{tabbing}

\noindent
where ${\sf DIV}$ is a process that immediately diverges. Although $P \refinedby[T] Q$ and $P \refinedby[F] Q$ (and vice-versa), $Q$ can continuously not accept $a$ and just perform $\tau$. The refinement $\refinedby[FD]$ allows such process comparisons. Formally, we consider both the failures and the divergences of a process $P$ as a pair: $(failures_\bot (P), divergences (P))$ and now define refinement in the failures/divergence model as follows: $P\refinedby[FD] Q \Leftrightarrow failures(Q) \subseteq failures(P) \wedge divergence(Q) \subseteq divergences(P)$. $failures_\bot (P)$ is defined as $failures(P) \cup \{ (s, X) \ | \ s \in divergences(P) \}$. That is, the set of traces leading to divergence are added to the existing set of stable failures to create an extended failures set.

\paragraph{Simplified Refinement Checking}

If both the specification and implementation are divergence free, then analysis only needs to demonstrate equivalence in the stable failures model. Indeed, for our verification purposes, both the specification and implementation are indeed divergence free, and therefore we only need check that our implementation meets the specification in the failures model.

\subsection{FDR}

FDR~\cite{fdr} is a refinement checker that can check various assertions about CSP processes written in \cspm\ (a machine readable version of CSP). FDR is used to analyse CSP models of systems and to test them against specifications (also written in CSP). FDR allows processes to be refinement checked via the three models shown in the previous section. FDR also supports assertions for checking deadlock freedom, divergence freedom, and whether a process definition is deterministic.

FDR can also check a system for determinism and divergence freedom. In the stable failures model, FDR ``considers a process $P$ to be deterministic providing no witness to non-determinism exists, where a witness consists of a trace {\em tr} and an event {\em a} such that $P$ can both accept and refuse {\em a} after {\em tr}.''~\cite{FDRManual}. In the failures-divergences model, the process must also be divergent free.

\subsection{Modelling Global State in CSP}
\label{sec:ModellingState}

All the ProcessJ runtime classes have member state variables. CSP does not have a global state space, only a global event space. It is therefore necessary to model such state in CSP using processes and events. In this section we present CSP that can be used to model general state that can be {\it set} and {\it read}. A state variable can be modelled in CSP via a process.  The process maintains the current value of the state variable, providing events to either {\em load} the current value of the variable, or {\em store} a new value of the variable.  We can define a generic process to model a state variable---$VARIABLE$---with alphabet $\alpha{}VARIABLE$ as follows:

\begin{tabbing}
=\===\= \kill
\>{\sf datatype} $Operations$ = $load\ \mid\ store$ \\
\>{\it VARIABLE}($var$, $val$) = \\
\>\>($var.load!val \then VARIABLE(var, val))$\\
\>\>$\extchoice$\\
\>\>$(var.store?val \then VARIABLE(var, val)$) \\
\>$\alpha{}VARIABLE(var, T) = \{ var.o.v\ |\ o \leftarrow Operations, v \leftarrow T \}$
\end{tabbing}

\noindent
$VARIABLE$ takes two parameters: $var$ is the channel used to communicate the variable with the environment.  $val$ is the current value of the variable.  In the alphabet $\alpha{}VARIABLE$, $val$ has type $T$, which is provided for the given variable by the system specifier.  

For example, each ProcessJ process has a {\em ready} and a {\it running} flag (stored as a field in a process), and we can define a channel for communicating these flags as:

\begin{tabbing}
=\===\= \kill
\>${\sf channel} \: ready : Processes.Operations.Bool$\\
\>${\sf channel} \: running : Processes.Operations.Bool$
\end{tabbing}
             % 4. CSP 
%!TeX root = ./Shared.tex
\section{Scheduling and Processes}
\label{sec:SchedulingAndProcesses}

Before we consider scheduling, we need to introduce the model of a ProcessJ {\it process}---or more precisely, the meta data associated with a process as well as the definition of a queue data structure in CSP.

\subsection{Processes}
\label{sec:processes2}

A process, in the context of this paper, is code the communicates; that is, either a reader or a writer. We implement these in Section~\ref{sec:ChannelsInCSP}. For each process the scheduling system requires metadata. Each process requires a monitor for mutual exclusion, a flag for running state, and a flag for the ready state. The fields are represented using the $VARIABLE$ process. We define a monitor process as:
\begin{tabbing}
=\===\= \kill
\>$MONITOR(claim, release)$ =\\
\>\>$claim?pid \then$\\
\>\>$release.pid \then$\\
\>\>$MONITOR(claim, release)$
\end{tabbing}
\noindent
where $claim$ and $release$ are channels communicating the {\it identifier} of the process claiming exclusive access to the metadata. $claim$ receives the identifier, and will only $release$ with the same identifier, ensuring exclusive access. We now define a $PROCESS$ as:
\begin{tabbing}
=\===\===\= \kill
\>$PROCESS(p)$ =\\ 
\>\>\hspace*{0.5cm}$VARIABLE(ready.p, true)$\\
\>\>$||| \ VARIABLE(running.p, false)$\\
\>\>$||| \ MONITOR(claim\_process.p, release\_process.p)$
\end{tabbing}
Running is initialised to false and ready to true. As we have multiple processes, we use $p$ to identify each process. We define $PROCESSES$ as the interleaving of all $PROCESS$ metadata object-processes:
\begin{tabbing}
=\===\===\= \kill
\>$PROCESSES = \ \underset{\mathclap{p \ \in Processes}}{|||} \ PROCESS(p)$
\end{tabbing}

\subsection{A Queue Process}

The scheduling system uses a queue of processes (the run queue), and we also require queues of processes that are waiting to use a shared channel end. We can implement a simple queue in CSP as follows:
\begin{tabbing}
=\===\===\= \kill
\>$QUEUE(enqueue, dequeue, size, q, CAP)$ =\\
\>\>\hspace*{0.5cm}$\# q < CAP\ \& \ enqueue?v \then QUEUE(enqueue, dequeue, q\ \hat{} <\!v\!>, CAP)$\\
\>\>$\extchoice \# q >$ 0 \& {\it dequeue}!{\sf head}($q$) $\then$ {\it QUEUE}({\it enqueue, dequeue}, {\sf tail}($q$), $CAP)$\\
\>\>$\extchoice \ size!(\# q) \then QUEUE(push, pop, size, q, CAP)$
\end{tabbing}
\noindent
where $\# q$ denotes the length of the sequence $q$. The queue can {\it enqueue} processes if it is not full; it can dequeue processes if it is not empty, and it can report its size. $CAP$ is the max size of the queue (this is needed to enable for FDR to verify the system---FDR cannot handle an unbounded queue). In an actual implementation we do not define a  max capacity. We can now completely describe the scheduling system in the next section.

\subsection{Scheduling}

In~\cite{ChalmersPedersen24} we fixed the problem with the divergence caused by not-ready processes in the run queue~\cite{PedersenChalmers}. We introduced a schedule manager to control the insertion of ready processes into the run queue.
\begin{tabbing}
=\===\===================\===\= \kill
\>$SCHEDULER$ = \\
\>\>$rqdequeue?p \then$ \>{\tt --} dequeue a process\\
\>\>$run.p \then$ \>{\tt --} run it\\
\>\>$yield.p \then$ \>{\tt --} wait for yield\\
\>\>$SCHEDULER$ \>{\tt --} recurse\\
$ $\\
\>$SCHEDULE\_MANAGER$ = \\
\>\>$schedule?pid \then$ \>{\tt --} wait for a schedule event\\
\>\>$rqenqueue!pid \then$ \>{\tt --} place the process in the run queue\\
\>\>$SCHEDULE\_MANAGER$ \>{\tt --} recurse\\
$ $\\
\>$SCHEDULERS(N)$ = \\
\>\>$(((\underset{\mathclap{n \ \in \{ 1 \dots N \} }}{\lvert\lvert\lvert} SCHEDULER) \interleave SCHEDULE\_MANAGER)$\\
\>\>\hspace*{0.5cm}$\underset{\mathclap{\alpha RUNQUEUE}}{||}$\\
\>\>\hspace*{0.5cm}$QUEUE(rqenque, rqdequeue, rqsize, <>, \# Processes)$\\
\>\>$)\ \backslash \ \alpha{}RUNQUEUE$ 
\end{tabbing}

In the original version of the scheduler, scheduling a process only required setting the {\it ready} field to true. With the new scheduler, a {\it running} field is introduced, requiring more considered scheduling operations. To schedule a process use the following procedure:

\begin{tabbing}
=\===\===\===\===\================\==\==\= \kill
\>$SCHEDULE(me, pid)$ = \\
\>\>$claim\_process.pid.me \then$ \>\>\>\>{\tt --} lock the process object\\
\>\>$ready.pid.load?r \then$ \>\>\>\>{\tt --} get the {\it ready} field\\
\>\>${\sf if}\ (r)\ {\sf then}$ \>\>\>\>{\tt --} {\bf if} (the process is ready) {\bf then}\\
\>\>\>$release\_process.pid.me \then$ \>\>\>{\tt --} \>release the lock\\
\>\>\>{\sf SKIP} \>\>\>{\tt --} \>and do nothing\\
\>\>{\sf else} \>\>\>\>{\tt --} {\bf else}\\
\>\>\>$ready.pid.store!{\sf true} \then$ \>\>\>{\tt --} \>set the process ready\\
\>\>\>$running.pid.load?r2 \then$ \>\>\>{\tt --} \>get the running status\\
\>\>\>${\sf if}\ (r2)\ {\sf then}$ \>\>\>{\tt --} \>{\bf if} (the proc. is running) {\bf then}\\
\>\>\>\>$release\_process.pid.me \then$ \>\>{\tt --} \>\>release the lock\\
\>\>\>\>{\sf SKIP}\>\>{\tt --} \>\>and do nothing\\
\>\>\>{\sf else} \>\>\>{\tt --} \>{\bf else}\\
\>\>\>\>$schedule.pid \then$ \>\>{\tt --} \>\>schedule the process\\
\>\>\>\>$release\_process.pid.me \then$ \>\>{\tt --} \>\>release the lock\\
\>\>\>\>{\sf SKIP}\>\>{\tt --} \>\>and do nothing
\end{tabbing}
Processes are descheduled when they yield.
\begin{tabbing}
=\===\===\===\===\================\==\==\= \kill
\>$YIELD(pid)$ = \\
\>\>$DESCHEDULE(pid);$ \>\>\>\> {\tt --} deschedule the process \\
\>\>$yield.pid \then$ \>\>\>\> {\tt --} yield to the scheduler \\
\>\>$run.pid \then$ \>\>\>\> {\tt --} wait to be scheduled \\
\>\>$running.pid.store!true \then$ {\sf SKIP} \>\>\>\> {\tt --} set running to true 
\end{tabbing}
Processes deschedule themselves, yield to the scheduler, and then wait for a scheduler to run them again. At this point, they set their running flag to true and continue from the point they yielded. Descheduling is defined as:
\begin{tabbing}
=\===\===\===\===\================\==\==\= \kill
\>$DESCHEDULE(pid)$ = \\
\>\>$claim\_process.pid.pid \then$ \>\>\>\> {\tt --} lock the process metadata\\
\>\>$running.pid.store.false \then$ \>\>\>\> {\tt --} set running to false\\
\>\>$ready.pid.load?r \then$ \>\>\>\> {\tt --} check if still ready to run\\
\>\>${\sf if}\ (r)\ {\sf then}$ \>\>\>\>{\tt --} {\bf if} (proc. ready to run) {\bf then}\\
\>\>\>$schedule.pid \then$ \>\>\>{\tt --} \>\> schedule the process\\
\>\>\>$release\_process.pid.pid \then$ \>\>\>{\tt --} \>\> unlock the  metadata\\
\>\>\>{\sf SKIP}\\
\>\>${\sf else}$ \>\>\>\> {\tt --} {\bf else} \\
\>\>\>$release\_process.pid.pid \then$ \>\>\>{\tt --} \>\> unlock the metadata\\
\>\>\>{\sf SKIP}
\end{tabbing}
The process metadata is locked, and the running flag set to false. A check is made on the ready flag, and if true it is scheduled (added to the run queue). Otherwise, nothing happens.

With all these processes in place, we can define a stand-alone scheduling process that works for any number of schedulers:

\begin{tabbing}
=\===\===\===\===\================\==\==\= \kill
	%$\settowidth\mylen{\small $\alpha{}SCHEDULING$}
\>$N\_SCHEDULER\_SYSTEM(N)$ =\\
\>\>\hspace*{6cm}$\mathclap{SCHEDULERS(N) \ _{\alpha{}SCHEDULING} || _{\alpha{}PROCESSES} \ PROCESSES}$  
\end{tabbing}

\begin{figure}
\centerline{\includegraphics[height=8cm]{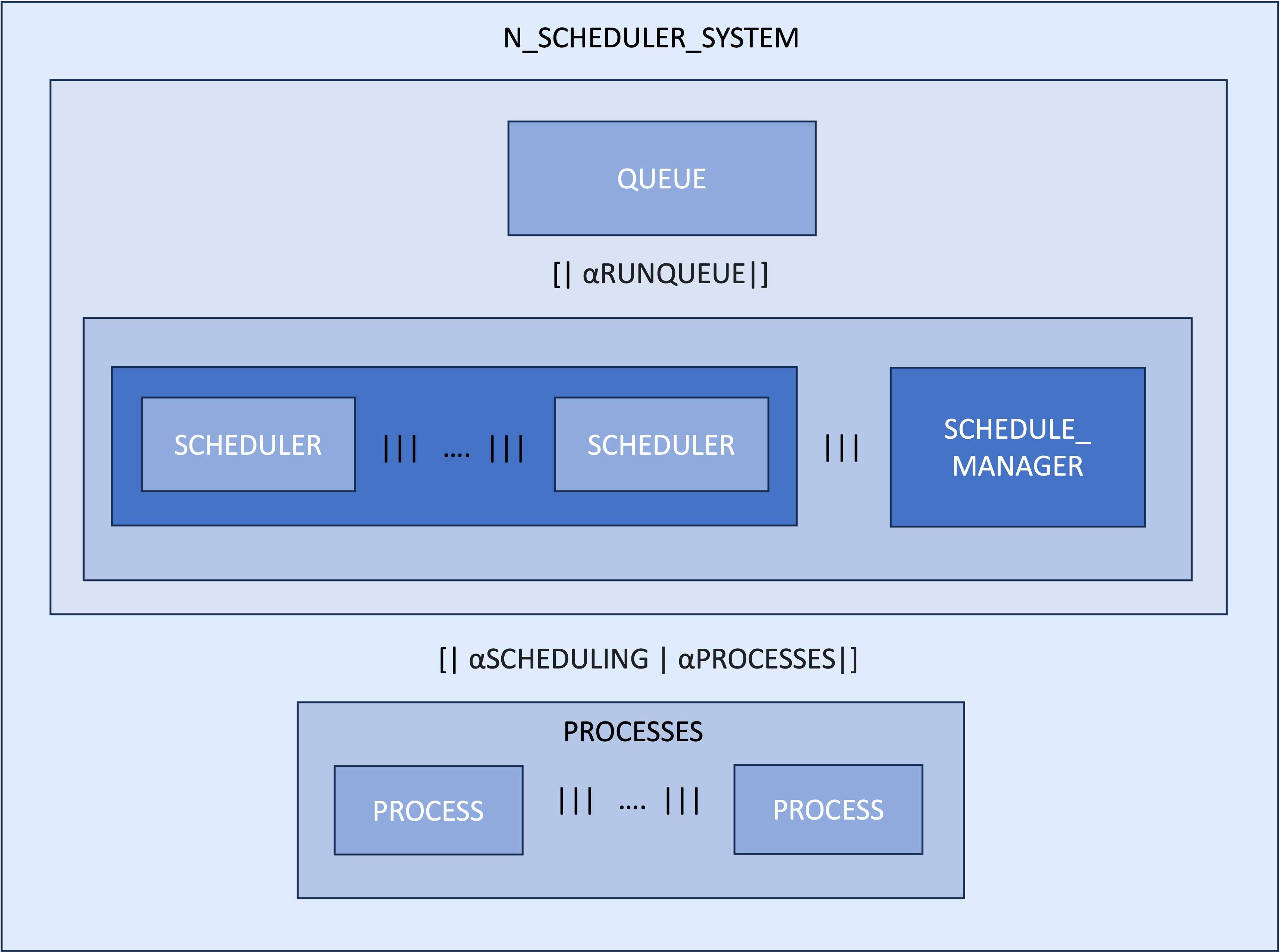}}
\caption{The $N\_SCHEDULER\_SYSTEM$ network.}
\label{fig:NSchedulerSystem}
\end{figure}

Figure~\ref{fig:NSchedulerSystem} shows the $N\_SCHEDULER\_SYSTEM$ process.

We now have the underlying scheduling system in place, and in the next section we present a the implementation of a shared channel in ProcessJ and its translation to CSP.
      % 5. Scheduling
%!TeX root = shared.tex
% Section 6
\section{Shared Channels in ProcessJ}
\label{sec:sharedChansInPJ}

In Section~\ref{sec:spec} we present the CSP for the {\bf specifications} of the behaviour of a shared channel. In this section we will develop the CSP for the {\it implementation} of the shared channel. This includes a number of steps to produce a CSP model starting with ProcessJ code and considering the generated Java code:
\begin{enumerate}
\item We need to consider the translation of both a channel {\it read} expression and a channel {\it write} statement in ProcessJ to Java.
\item We need to translate this generated Java code into CSP.
\item We need CSP code for the ProcessJ runtime implementation of the shared channel (written in Java), this includes functionality to claim and unclaim channel ends amongst other things.
\end{enumerate}

In the following section we start with an example of two ProcessJ programs that use shared channels; one with a shared reading end and one with a shared writing end. 

\subsection{ProcessJ Example of Shared Ends}
\label{sec:PJexSharedEnds}

Figure~\ref{fig:sharedEnds} shows two different ProcessJ programs. The left side illustrates the use of a channel with a \textbf{shared writing} end: we run two instances of $f$ that share the writing end of a channels \textit{c} concurrently with a single reader. The reader reads from the channel twice and prints out either {\tt 42 43} or {\tt 43 42} depending on which writer uses the channel first. The right side illustrates the use of a channel with a \textbf{shared reading} end: we run one writer (that writes the values {\tt 42} and then {\tt 43} on the channel) along with two readers. Each reader will read from the shared channel once and print their values. Depending on the scheduling, the result is similarly either {\tt 42 43} or {\tt 43 42}.

\begin{figure}
\begin{center}
\noindent
\begin{tabular}{|l|l|}\hline
\multicolumn{1}{|c|}{\textbf{Shared Writing End}} & \multicolumn{1}{|c|}{\textbf{Shared Reading End}}\\\hline
\begin{minipage}{7cm}
$ $\\
{\bf void} \textit{f}(\textbf{shared} \textbf{chan}$<$\textbf{int}$>$.\textbf{write} \textit{out},\\
\hspace*{32pt}\textbf{int} \textit{val}) \{\\
\hspace*{15pt}\textit{out}.\textbf{write}(\textit{val});\\
\}\\
$ $\\
\textbf{void} \textit{g}(\textbf{chan}$<$\textbf{int}$>$.\textbf{read} \textit{in})\{\\
\hspace*{15pt}\textbf{int} \textit{x};\\
\hspace*{15pt}\textit{x} = \textit{in}.\textbf{read}();\\
\hspace*{15pt}\textit{println}(x);\\
\hspace*{15pt}\textit{x} = \textit{in}.\textbf{read}();\\  
\hspace*{15pt}\textit{println}(x);\\  
\}\\
$ $\\  
\textbf{void}\textit{main}(\textbf{string} \textit{args}[ ]) \{\\
\hspace*{15pt}\textbf{shared write} \textbf{chan}$<$\textbf{int}$>$ \textit{c};\\
\hspace*{15pt}\textbf{par} \{\\
\hspace*{30pt}\textit{f}(\textit{c}.\textbf{write}, 42);\\
\hspace*{30pt}\textit{f}(\textit{c}.\textbf{write}, 43);\\
\hspace*{30pt}\textit{g}(\textit{c}.\textbf{read});\\ 
\hspace*{15pt}\}\\
\}\\
\end{minipage}
&
\begin{minipage}{7cm}
$ $\\
{\bf void} \textit{f}(\textbf{chan}$<$\textbf{int}$>$.\textbf{write} \textit{out}, \textbf{int} \textit{val}) \{\\
\hspace*{15pt}\textit{out}.\textbf{write}(\textit{val});\\
\hspace*{15pt}\textit{out}.\textbf{write}(\textit{val+1});\\
\}\\
$ $\\
\textbf{void} \textit{g}(\textbf{shared} \textbf{chan}$<$\textbf{int}$>$.\textbf{read} \textit{in}) \{\\
\hspace*{15pt}\textbf{int} \textit{x};\\
\hspace*{15pt}\textit{x} = \textit{in}.\textbf{read}();\\  
\hspace*{15pt}\textit{println}(\textit{x});\\
\}\\
$ $\\  
\textbf{void} \textit{main}(\textbf{string} \textit{args}[ ]) \{\\
\hspace*{15pt}\textbf{shared read} \textbf{chan}$<$\textbf{int}$>$ \textit{c};\\
\hspace*{15pt}\textbf{par} \{\\
\hspace*{30pt}\textit{f}(\textit{c}.\textbf{write}, 42);\\
\hspace*{30pt}\textit{g}(\textit{c}.\textbf{read});\\ \hspace*{30pt}\textit{g}(\textit{c}.\textbf{read});\\ 
\hspace*{15pt}\}\\
\}\\
$ $\\
$ $\\
\end{minipage}\\\hline
\multicolumn{2}{l}{}\\
\end{tabular}
\end{center}
\caption{Example ProcessJ shared reading (one-to-many) and shared writing (many-to-one) channels.}
\label{fig:sharedEnds}
\end{figure}

Channels can also have both shared reading and writing ends. A shared channel is \textbf{not} a CSP multi-way synchronisation, which requires all involved processes to engage on the same event. With shared channels, only one reader and one writer are engaged per channel communication.

In order to ensure that only one writer/reader will be accessing the channel, we require a mutual exclusion mechanism. In ProcessJ, a claim value and process queue is used. Next, we consider how to translate the processJ communication primitives into Java.

\subsection{Translating {\it write}() to Java on a Shared Channel Write End}
\label{sec:writeToJava}

As the first step we consider the translation of {\it write} in ProcessJ to Java. A ProcessJ {\it write} statement on a \textit{non-shared} channel end is:

\begin{tabbing}
---\=\kill
\>{\bf chan}$<\cdots>$ $c$;\\
\>$\vdots$\\
\>\textit{c}.\textbf{write}(\textit{val})
\end{tabbing}
\noindent
This code translates to the following Java code:
\begin{tabbing}
---\=\kill
\>\textit{$\overline{c}$}.\textit{write}(\textbf{this}, \textit{val});\\
\>\textbf{this}.\textit{runLabel} = 2; // representing L$_2$\\
\>\textit{yield}();\\
\>L$_2$:
\end{tabbing}

\noindent
where $\overline{c}$ is a Java object reference to a channel. If a channel end is shared, access to it must be protected. The simplest approach is to claim the channel before the code above gets executed and release it again after. In order to avoid busy waiting for processes that fail the claim, we equip a shared channel end with a queue of processes waiting to use the channel end. Therefore, if a process fails to claim, it gets placed at the end of the queue. It it worth noting that a channel with shared ends is no different in its basic construction than a non-shared channel except access to it is controlled with a claim value and a queue of processes awaiting access. If we repeat the ProcessJ code from above with a channel where the writing end is shared we have:

\begin{tabbing}
---\=\kill
\>{\bf shared write chan}$<\cdots> c$;\\
\>$\vdots$\\
\>$c$.{\it write}({\it val});
\end{tabbing}

\noindent
The generated Java code with a claim looks like this (code in outlined boxes comes from the non-shared version):

\begin{tabbing}
---\=---\=\kill
\>\textbf{if} (!\textit{$\overline{c}$}.\textit{claimWrite}(\textbf{this})) \{\\
\>\>\textbf{this}.\textit{runLabel} = 1; // representing L$_1$\\
\>\>\textit{yield}();\\
\>\}\\ 
\>L$_1$:
\end{tabbing}

\framebox{\begin{varwidth}{\linewidth}
\begin{tabbing}
---\=---\=\kill
\textit{$\overline{c}$}.\textit{write}(\textbf{this}, \textit{val});\\
\textbf{this}.\textit{runLabel} = 2; // representing L$_2$\\
\textit{yield}();\\
L$_2$:
\end{tabbing}
\end{varwidth}}
\begin{tabbing}
---\=---\=\kill
\>\textit{$\overline{c}$}.\textit{unclaimWrite}();
\end{tabbing}

\noindent
where $\overline{c}$ is a Java reference to a channel object, and {\bf this} is a reference to the process object ({\it PJProcess}). Before the write can commence on the shared channel end, the process must first make a claim on the writing end of the shared channel. This is done in the synchronized \textit{claimWrite}() method. The claim succeeds if no other process holds a claim or if the process already has a claim on the channel end. If a different process already claimed the channel end, the process will be placed in a queue, set not-ready to run and \textit{false} will be returned. If the claim fails, the process yields, and when re-awoken (continuing at L$_1$), the write can proceed. The process was set ready to run by a different process that held the lock on the channel end. This happens in \textit{unclaimWrite}(). After this, the process again yields in order for the communication to remain synchronous (the writing process gets re-started by the reader eventually) and continues at L$_2$, after which the channel end is unclaimed and ready to be claimed by another writer. When a process unclaims a channel end, the queue of waiting processes is popped (if not empty) and the removed process is set ready to run and registered as holding the claim to the channel end.

The channel end claims on a shared channel uses the field \textit{writerclaim} for the shared write channels, and \textit{readclaim} for the shared read channels. This field is set and read in synchronized methods only. Both the \textit{claimWrite} and \textit{unclaimWrite} are synchronous methods. This means that a different lock is taken out on the entire channel object while these two methods run. We will model this in the implementation as well. Let us look at the \textit{PJMany2OneChannel} class:

\noindent
\begin{footnotesize}
\begin{tabbing}
=\===\===\===\===\================\==\==\= \kill
\>\textbf{public} \textbf{class} \textit{PJMany2OneChannel}$<$\textit{T}$>$ \textbf{extends} \textit{PJOne2OneChannel}$<$\textit{T}$>$ \{\\ 
\>\>\textbf{protected} \textit{PJProcess} \textit{writeclaim} = \textbf{null};\\
$ $\\
\>\>\textbf{protected} \textit{Queue}$<$\textit{PJProcess}$>$ \textit{writeQueue} = \textbf{new} \textit{LinkedList}$<>$();\\
$ $\\
\>\>\textbf{public} \textbf{synchronized} \textbf{boolean} \textit{claimWrite}(\textit{PJProcess} \textit{p}) \{\\
\>\>\>\textbf{if} (\textit{writeclaim} == \textbf{null} $\mid\mid$ \textit{writeclaim} == \textit{p}) \{\\
\>\>\>\>\textit{writeclaim} = \textit{p}; // this channel's write is claimed by $p$. \\
\>\>\>\>\textbf{return} \textbf{true};\\
\>\>\>\}\textbf{else} \{\\
\>\>\>\>\textit{p.setNotReady}(); // set $p$ not ready to run.\\
\>\>\>\>\textit{writeQueue}.\textit{add}(\textit{p}); // add $p$ to the queue of waiting writers.\\
\>\>\>\}\\
\>\>\>\textbf{return} \textbf{false};\\
\>\>\}\\
$ $\\    
\>\>\textbf{public} \textbf{synchronized} \textbf{void} \textit{unclaimWrite}() \{\\
\>\>\>if (\textit{writeQueue.isEmpty}()) \{\\
\>\>\>\>\textit{writeclaim} = \textbf{null}; // no one wants the channel's writing end right now.\\
\>\>\>\} \textbf{else} \{\\
\>\>\>\>\textit{PJProcess} \textit{p} = \textit{writeQueue.remove}(); // get the next waiting writer.\\
\>\>\>\>\textit{writeclaim} = \textit{p}; // this channel's write end is claimed by $p$.\\
\>\>\>\>\textit{schedule}($p$); // schedule $p$.\\ 
\>\>\>\}\\ 
\>\>\}\\
\>\}
\end{tabbing}
\end{footnotesize}

The field {\it writeclaim} holds a reference to the process that currently holds the lock on the channel or {\tt null} when no claim exists. The {\it writeQueue} holds a list of other writers that wish to write to the channel.

The two methods \textit{claimWrite}() and \textit{unclaimWrite}() are used for claiming and unclaiming a channel. Let us consider them in greater detail in turn. A call to \textit{claimWrite}() with a process object reference checks if the channel is currently claimed or otherwise. If true, \textit{writeclaim} is updated to the process' reference and \textit{true} is returned. If another process has a claim of the channel, the waiting process is set not-ready and placed at the back of the queue and \textit{false} is returned.

\textit{unclaimWrite}() will set the {\it writeclaim} field back to \textit{null} if no other processes are waiting in the queue. If one is, the first will be removed from the queue and its reference is placed in \textit{writeclaim}, \textbf{and} that process will be set ready. It should be clear now why the second check (\textit{writeclaim} == $p$) is necessary in the \textit{claimWrite}() method.

To model this in CSP we need:

\begin{itemize}
\item A variable \textit{writeclaim} holding a process ID.
\item A queue of process IDs.
\item The code for \textit{claimWrite}() and \textit{unclaimWrite}()
\end{itemize}
in addition, we need the state and methods from the \textit{PJOne2OneChannel} class, which are as follows:
\begin{itemize}
\item The \textit{reader} and \textit{writer} variables representing the reader and the writer of the channel.
\item Code for \textit{isReadyToRead}(), \textit{read}() and \textit{write}().
\end{itemize}
but also the data from the \textit{PJChannel} class:
\begin{itemize}
\item The actual \textit{data} on the channel.
\end{itemize}

\noindent
For completeness, let us just list the code for \textit{PJOne2OneChannel} and \textit{PJChannel} here:

\noindent
\begin{tabbing}
\hspace*{8cm} \= \kill
{\bf public} $PJOne2OneChannel\!\!<\!\!T\!\!>$ {\bf extends} $PJChannel\!\!<\!\!T\!\!>$ \{\\
\ind2  $PJProcess\ reader = null$; \\
\ind2  $PJProcess\ writer = null$; \\
\\
\ind2  {\bf boolean} $isReadyToRead$($PJProcess\ p$) \{ \\
\ind3    {\bf if} ($writer\ != null$) \{  \>// {\bf if} (a writer is present) {\bf then}\\
\ind4      {\bf return} {\bf true};         \>// \quad{}{\bf return} true\\
\ind3    \} \textbf{else} \{              \>// {\bf else}\\
\ind4      $reader\ = p$;                   \>// \quad{}register {\it p} as reader\\
\ind4      $reader.setNotReady()$;          \>// \quad{}set reader not ready\\
\ind4      {\bf return false};       \>// \quad{}{\bf return} false\\
\ind3\} \\
\ind2\} \\
\\
\ind2{\bf void} $write$($PJProcess\ p,\ T\ item$) \{ \\
\ind3  $data = item;$                \>// set data on channel\\
\ind3  $writer = p$;                 \>// register the writer\\
\ind3  $writer.setNotReady()$;       \>// set writer not ready \\
\ind3  {\bf if} ($reader\ != null$)  \>// {\bf if} (a reader is there) {\bf then}\\
\ind4  $schedule(this, p)$;                \>// \quad{}schedule it\\ 
\ind2\} \\
\\
\ind2$T\ read$($PJProcess\ p$) \{ \\
\ind3  $schedule(this, writer)$;          \>// schedule the writer\\ 
\ind3  $writer = null$;             \>// clear writer field\\
\ind3  $reader = null$;\>// clear reader field\\
\ind3  {\bf return} $data$; \>// {\bf return} the data \\
\ind2\} \\
\ind2$\vdots$  \\
\ind1\} 
\end{tabbing}
\noindent
and the parts of \textit{PJChannel} that are important:\\

\noindent
\ind1{\bf public class} $PJChannel\!<\!T\!> \{$\\
\ind2{\bf protected} $T\ data$;\\    
\ind2$\cdots$\\
\ind1\}\\

The CSP code for \textit{read}() and \textit{write}() was previously developed and used in~\cite{PedersenChalmers}. Since we are considering a shared-write (many-to-one) channel, the read expression (\textit{c}.\textbf{read}()) will be exactly as it was in~\cite{PedersenChalmers}, namely:

\begin{tabbing}
---\=---\=\kill
\>{\bf if} ($!\overline{c}.isReadyToRead$()) \{\\ 
\>\>{\bf this}.{\it runLabel} = 1; // representing L$_1$\\
\>\>$yield$();\\
\>\}\\
\>L$_1$:\\
\>$var = \overline{c}.read$();\\
\>{\bf this}.{\it runLabel} = $2$; // representing L$_2$\\
\>$yield$();\\
\>$L_2$:
\end{tabbing}

\noindent
Again, the label numbers 1 and 2 are arbitrarily chosen here. Before we develop the CSP, let us first consider shared read channels in the next section.

\subsection{Translating \textit{read}() to Java on a Shared Channel Read End}

Like before, let us start with the read on a \textit{non-shared} channel read end:

\begin{tabbing}
---\=---\=\kill
\>{\bf chan}$<\cdots>$ $c$;\\
\>$\vdots$\\
\>\textit{var} = \textit{c}.\textbf{read}()
\end{tabbing}

\noindent
We saw the code that this expression generates for a non-shared channel end at the end of the previous section. The approach is very similar to the shared writing end (where the channel is defined as {\bf shared read chan}$<\cdots>$): we claim a lock before and release after, and we maintain a queue of readers that failed to obtain the lock\footnote{The actual code for the {\it claimRead}() and {\it unclaimRead}() can be found in the GitHub repository along with all the CSP scripts used in this paper.}. This generated Java code looks like this:

\begin{tabbing}
---\=---\=\kill
\>\textbf{if} (!\textit{$\overline{c}$}.\textit{claimRead}(\textbf{this})) \{ \\
\>\>\textbf{this}.\textit{runLabel} = 1; // representing L$_1$\\
\>\>\textit{yield}();\\
\>\} \\
\>L$_1$;\\[-0.75cm]
\end{tabbing}
\,\,\,\,\,\framebox{\begin{varwidth}{\linewidth}
\begin{tabbing}
---\=---\=\kill
\textbf{if} (!\textit{$\overline{c}$}.\textit{isReadyToRead}(\textbf{this})) \{\hspace*{1cm} \\
\>\textbf{this}.\textit{runLabel} = 2; // representing L$_2$\\
\>\textit{yield}();\\
\}\\
L$_2$;\\\\
\textit{x} = \textit{$\overline{c}$}.\textit{read}(\textbf{this});\\
\textbf{this}.\textit{runLabel} = 3; // representing L$_3$
\end{tabbing}
\end{varwidth}}
\begin{tabbing}
---\=\kill\\[-0.75cm]
\>\textit{$\overline{c}$}.\textit{unclaimRead}();\\[-0.75cm]
\end{tabbing}
\,\,\,\,\,\framebox{\begin{varwidth}{\linewidth}
\begin{tabbing}
---\=---\kill
\textit{yield}();\hspace*{4.45cm}\\
L$_3$;
\end{tabbing}
\end{varwidth}}
$ $\\
$ $\\

Note, the placement of the \textit{unclaimRead}() before the \textit{yield}(). The yield is a courtesy yield\footnote{We do this to create some kind of fairness.} only and could be removed, so it is OK to unclaim the channel end before performing the yield.

The \textit{PJOne2ManyChannel} class is very similar to the \textit{PJMany2OneChannel} that we showed earlier; rather than \textit{claimWrite}() and \textit{unclaimWrite}() it has \textit{claimRead}() and \textit{unclaimRead}(). Rather than a \textit{writeQueue} it has a \textit{readQueue}, and instead of the \textit{writeclaim} field it has a \textit{readclaim} field. The implementation of the claim and unclaim method are exactly the same as for the many-to-one channel.

We know that we can reuse the \textit{read}() and \textit{write}() code from~\cite{PedersenChalmers} as well as code generated for the writes to non-shared ends as reads from non-shared ends. For shared ends, we need to book-end the code it with the claiming and unclaiming of the channel end as just explained and illustrated. In the next section we present the needed CSP for the new variables/fields and the required queue.

 % 6. Shared Chans in PJ
% !TeX root = ./shared.tex
% Section 7
\section{CSP Model of the ProcessJ shared channels}
\label{sec:imp}

The next step is to translate the generated Java code into CSP and then perform the verification against the specifications from Section~\ref{sec:spec}. However, we start with the ProcessJ channel implementation in CSP.

\subsection{Channels in CSP}
\label{sec:ChannelsInCSP}

As we are implementing channels with shared ends (reading and writing), we define a {\it Processes} type based on the tests being undertaken. For example, for a one-to-many channel with one writer sending to two readers we have:

\begin{tabbing}
=\===\===\===\===\================\==\==\= \kill
\>{\tt --} all processes used by the model\\
\>{\sf datatype} {\it Nullable\_Processes = NULL} \; | \; {\it R1} \; | \; {\it R2} \; | \; {\it W}\\
\> {\tt --} the processes that can be scheduled\\
\>{\sf subtype} {\it Processes = R1} \; | \; {\it R2} \; | \; {\it W}\\
\> {\tt --} the processes reading from the channel\\
\>{\sf subtype} {\it Reading\_Processes = R1} \; | \; {\it R2}\\
\> {\tt --} the processes writing to the channel\\
\>{\sf subtype} {\it Writing\_Processes = W}\\
\>{\tt --} the reading processes with null\\
\>{\sf subtype} {\it Nullable\_Reading\_Processes = NULL} \; | \; {\it R1} \; | \; {\it R2}\\
\>{\tt --} the writing processes with null\\
\>{\sf subtype} {\it Nullable\_Writing\_Processes = NULL} \; | \; {\it W}
\end{tabbing}

\begin{figure}
	\centering
	\includegraphics[width=0.3\textwidth]{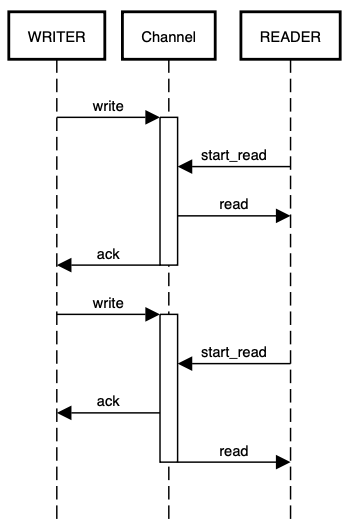}
	\hspace*{0.1\textwidth}
	\includegraphics[width=0.3\textwidth]{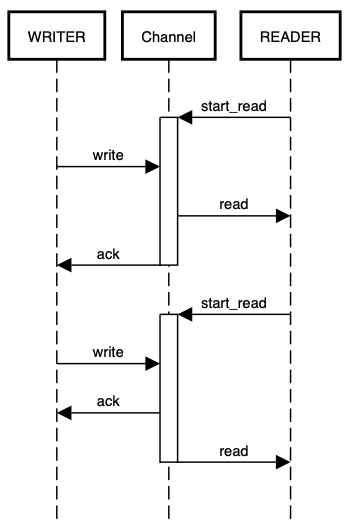}
	\caption{Four interactions of channels.}
	\label{fig:interactions}
\end{figure}

A channel communication in ProcessJ is not instantaneous as was described in Section~\ref{sec:PJexSharedEnds}. Both the writer and the reader on a channel start their respective interaction, wait for synchronisation, and then complete. We consider these two operations (reading and writing) to have a starting event and an ending event. Due to the reader and writer effectively interleaving except at the channel synchronisation point, their are four possible interactions that can occur on a channel as presented in Figure~\ref{fig:interactions}.
We therefore must define these four events as channels like so:
\begin{tabbing}
=\===\===\===\===\================\==\==\= \kill
\>{\sf channel} {\it read, write$: $Channels.Processes.Values}\\
\>{\sf channel} \textit{start\_read}, \textit{ack}: \textit{Channels}.\textit{Processes}
\end{tabbing}
A {\it read} and {\it write} communicate on a {\it channel}. We identify the {\it process} performing the action, and need to know the {\it value} communicated. We therefore must define the types $Channels$ and $Values$. Our system only has one channel. Two values are sufficient to demonstrate the correct message transmission of a channel.
\begin{tabbing}
=\===\===\===\===\================\==\==\= \kill
\>{\sf datatype} {\it Channels = C1} \\
\>{\sf datatype} {\it Values = A} \ | \ {\it B}
\end{tabbing}
In Section~\ref{sec:SchedulingAndProcesses} we introduced the {\it VARIABLE} and the {\it MONITOR} processes. These are needed in our channel implementation as well. From~\cite{PedersenChalmers} we can almost reuse the {\it CHANNEL} process; all we need extra is a monitor. A channel has a single monitor as well as three fields: a reference to the reader, a reference to the writer, and the data being sent; we can use three {\it VARIABLE} processes to implement these fields: 
\begin{tabbing}
=\===\===\===\===\================\==\==\= \kill
\>{\sf channel} {\it writer, reader : Channels.Operations.Nullable\_Processes}\\
\>{\sf channel} {\it data : Channels.Operations.Values}\\
\>{\sf channel} {\it channel\_claim, channel\_release :Channels.Processes}\\
$ $\\
\>{\it CHANNEL}$(chan)$ =\\  
\>\>\hspace*{0.5cm}{\it VARIABLE}$(writer.chan, NULL)$\\ 
\>\>$\interleave$ {\it VARIABLE}$(reader.chan, NULL)$ \\
\>\>$\interleave$ {\it VARIABLE}$(data.chan, A)$\\
\>\>$\interleave$ {\it MONITOR}$(channel\_claim.chan, channel\_release.chan)$
\end{tabbing}
\noindent
We also reuse the code for \textit{read}() and \textit{write}():
\begin{tabbing}
=\===\===\===\===\================\==\==\= \kill
\>{\it READ}$(chan)$ = \\
\>\>$writer.chan.load?w \then\hspace*{1.5cm}$ \>\>\>\> \mbox{{\tt --} get $writer$.}\\
\>\>{\sf if} ($w$ == $NULL$)\ {\sf then}\> \\
\>\>\hspace*{15pt}{\sf DIV} \>\>\>\> \mbox{{\tt --} this never happens}\\ 
\>\>{\sf else}\> \\
\>\>$\hspace*{15pt}ready.w.store!true \then$      \>\>\>\>\mbox{{\tt --} $writer.setReady$();}\\
\>\>$\hspace*{15pt}writer.chan.store!NULL \then$  \>\>\>\> \mbox{{\tt --} $writer = null$;}\\
\>\>$\hspace*{15pt}reader.chan.store!NULL \then$  \>\>\>\> \mbox{{\tt --} $reader = null$;}\\
\>\>\hspace*{15pt}{\sf SKIP}\> \\
$ $\\
\>{\it WRITE}$(pid, chan, item)$ =\\
\>\>$data.chan.store!item \then$ \>\>\>\>  \mbox{{\tt --} $data$ = $item$;}\\        
\>\>$writer.chan.store!pid \then$    \>\>\>\>  \mbox{{\tt --} $writer$ = $pid$;}\\         
\>\>$ready.pid.store!false \then$    \>\>\>\>  \mbox{{\tt --} $writer.setNotReady$();}\\        
\>\>$reader.chan.load?v \then$      \>\>\>\>  \mbox{{\tt --} get $reader$} \\
\>\>${\sf if}\ (v != NULL)\ {\sf then}$ \>\>\>\>  \mbox{{\tt --} i{\bf f} ($reader$ != $null$)} {\bf then}\\
\>\>$\hspace*{15pt}ready.v.store!true \then {\sf SKIP}\qquad$  \>\>\>\>  \mbox{{\tt --} \quad{}$schedule(this, reader)$;}\\ 
\>\>${\sf else}$ \\
\>\>$\hspace*{15pt}{\sf SKIP}$ \> 
\end{tabbing}
\noindent
To interact with the {\it MONITOR} and {\it QUEUE} processes we define the following channels:
\begin{footnotesize}
\begin{tabbing}
=\===\===\===\===\================\==\==\= \kill
\>{\sf channel} {\it write\_end\_enqueue, write\_end\_dequeue : Channels.Writing\_Processes}\\
\>{\sf channel} {\it read\_end\_enqueue, read\_end\_dequeue : Channels.Reading\_Processes}\\
\>{\sf channel} {\it write\_queue\_size : Channels.\{0..(card(Writing\_Processes) - 1)\}}\\
\>{\sf channel} {\it read\_queue\_size : Channels.\{0..(card(Reading\_Processes) - 1)\}}\\
\>{\sf channel} {\it readclaim : Channels.Operations.Nullable\_Reading\_Processes}\\
\>{\sf channel} {\it writeclaim : Channels.Operations.Nullable\_Writing\_Processes}
\end{tabbing}
\end{footnotesize}
{\it *\_end\_enqueue}\, and {\it *\_end\_dequeue}\, are used to add or remove processes to the wait queue on a shared channel end. The {\it *\_queue\_size} channels allow getting the current size of the wait queues. Finally, {\it readclaim} and {\it writeclaim} allow getting and setting the current process claiming the shared end.
\begin{figure}
\centerline{\includegraphics[height=5cm]{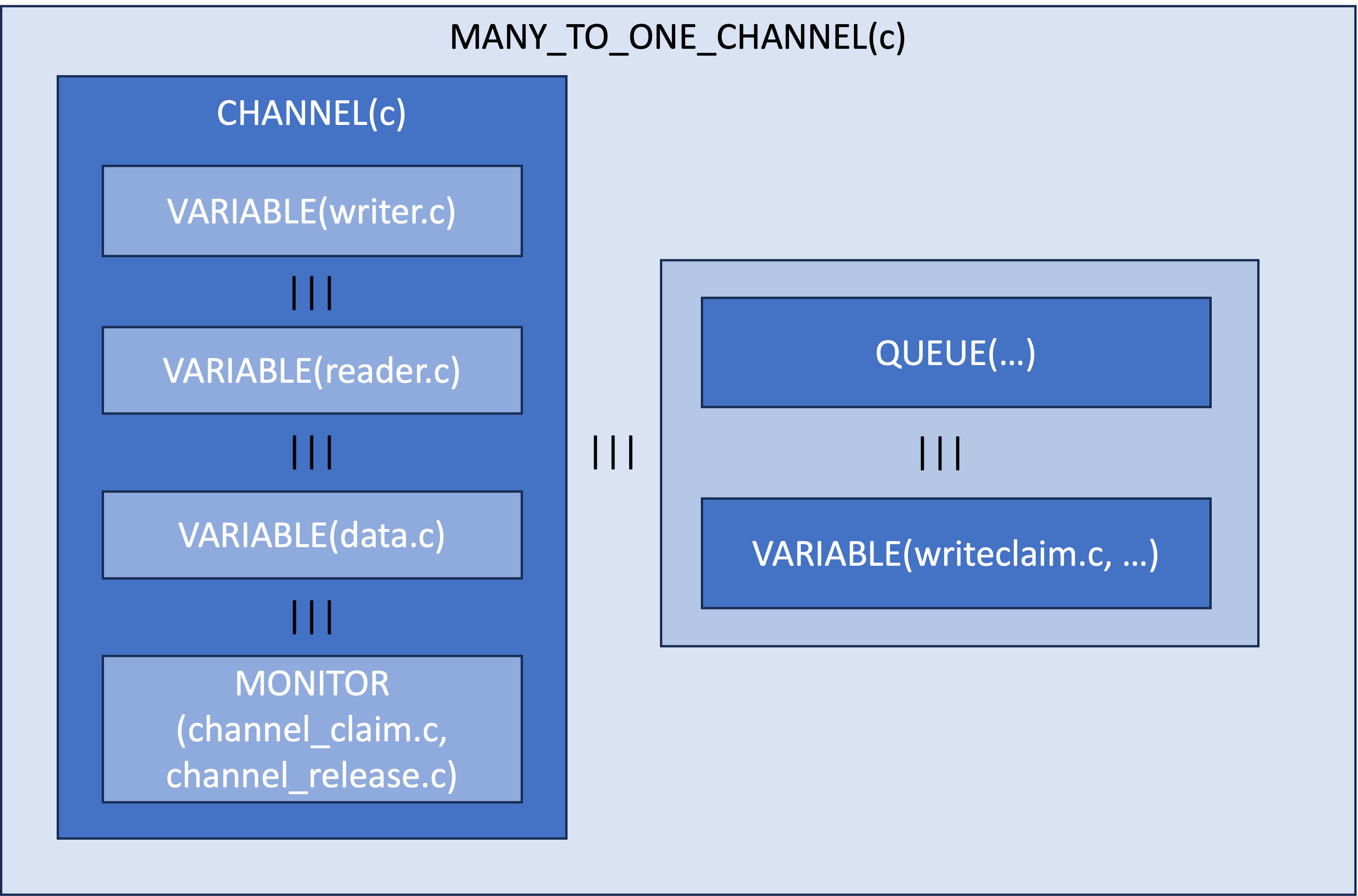}}
\caption{The {\it MANY\_TO\_ONE\_CHANNEL} process.}
\label{fig:many2one}
\end{figure}
We create a many-to-one channel process by interleaving a regular channel, a queue of writers, and a variable for the field \textit{writeclaim}:
\begin{tabbing}
=\===\===\===\===\================\==\==\= \kill
\>{\it MANY\_TO\_ONE\_CHANNEL}$(c) =$\\
\>\>\hspace*{0.5cm}{\it CHANNEL}$(c)$\\
\>\>$\interleave$ {\it QUEUE}$(end\_enqueue.c.write\_end, end\_dequeue.c.write\_end,$\\
\hspace*{3.2cm}$queue\_size.c.write\_end, <\ >)$\\
\>\>$\interleave$ {\it VARIABLE}$(writeclaim.c, NULL)$
\end{tabbing}

\noindent
Figure~\ref{fig:many2one} illustrates the {\it MANY\_TO\_ONE} channel.

For the \textit{ONE\_TO\_MANY\_CHANNEL} process, the first two parameters of the \textit{QUEUE} process simply changes to \textit{end\_enqueue.c.read\_end, end\_dequeue.c.read\_end}, and the \textit{VARIABLE}'s first parameter is now the \textit{readclaim} channel instead. For the \textit{MANY\_TO\_MANY\_CHANNEL} process, we need two queues (one for readers and one for writers) and two variables (one for the \textit{readclaim} and one for the \textit{writeclaim} fields). The complete code is available on the GitHub page listed for this paper in the introduction.

\subsection{Claiming and Unclaiming a Channel}

We now implement the {\it CLAIM\_WRITE} and {\it UNCLAIM\_WRITE} as follows:

\begin{tabbing}
=\===\===\===\===\=================\==\==\= \kill
\>{\it CLAIM\_WRITE}$(pid, chan) =$\\
\>\>$channel\_claim.chan!pid \then$ \>\>\>\> {\tt --} claim the channel\\
\>\>$writeclaim.chan.load?wc \then$  \>\>\>\> {\tt --} read the {\it writeclaim} field\\
\>\>{\sf if} $(wc == NULL\ {\sf or}\ wc == pid)\ {\sf then} $ \>\>\>\> {\tt --} \textbf{if} (\textit{writeclaim} == \textbf{null} $\mid\mid$\\
\>\>\>\>\>\>\>\>\hspace*{0.33cm}\textit{writeclaim} == \textit{p}) {\bf then}\\
\>\>\>$writeclaim.chan.store!pid \then$ \>\>\> {\tt --} \>\>\textit{writeclaim = p};\\
\>\>\>$channel\_release.chan.pid \then$ \>\>\> {\tt --}\>\>release the channel\\
\>\>\>{\sf SKIP}\\
\>\>{\sf else} \>\>\>\>{\tt --} {\bf else}\\
\>\>\>$ready.pid.store!{\sf false} \then$ \>\>\> {\tt --}\>\>\textit{p.setNotReady}();\\
\>\>\>$write\_end\_enqueue.chan.pid \then$ \>\>\> {\tt --} \>\>$writeQueue.add(p);$\\
\>\>\>$channel\_release.chan.pid \then$ \>\>\> {\tt --}\>\> release the channel\\
\>\>\>{\it YIELD}$(pid)$ \>\>\> {\tt --} \>\>$yield();$\\
$ $\\
\>{\it UNCLAIM\_WRITE}$(pid, chan)$ =\\
\>\>$channel\_claim.chan!pid \then$ \>\>\>\> {\tt --} claim the channel \\
\>\>$queue\_size.chan.write\_end?s \then$ \>\>\>\> {\tt --} $s = writeQueue.size();$\\
\>\>{\sf if} $(s == 0)$ {\sf then} \>\>\>\> {\tt --} {\bf if} ($s == 0)$ {\bf then}\\
\>\>\>$writeclaim.chan.store!NULL\ \then$ \>\>\> {\tt --} \>\>$writeclaim = null;$\\
\>\>\>$channel\_release.chan.pid \then$  \>\>\> {\tt --} \>\>release the channel\\
\>\>\>{\sf SKIP}\\
\>\>{\sf else} \>\>\>\> {\tt --} \textbf{else}\\
\>\>\>$write\_end\_dequeue.chan?p \then$ \>\>\> {\tt --} \>\>$p = writeQueue.pop();$\\
\>\>\>$writeclaim.chan.store!p \then$ \>\>\> {\tt --}\>\>$writeclaim = p;$\\
\>\>\>{\it SCHEDULE}$(pid, p);$ \>\>\> {\tt --}\>\>$schedule(p);$\\ 
\>\>\>$channel\_release.chan.pid \then$ \>\>\> {\tt --}\>\>release the channel\\
\>\>\>{\sf SKIP} 
\end{tabbing}
\noindent
{\it SCHEDULE} schedules the process by adding it to the run queue and setting its {\it ready} field to {\it true} if false. The definitions for {\it CLAIM\_READ} and {\it UNCLAIM\_READ} are similar as expected.

We also define the CSP for the ProcessJ channel-write statement and channel-read expression; we call these \textit{RE\-STRIC\-TED\_\-PRO\-CESS\_\-WRI\-TER} for a non-shared writing end, \textit{RE\-STRIC\-TED\_PRO\-CESS\_SHAR\-ED\_WRIT\-ER} for a share writing end, \textit{RE\-STRIC\-TED\_PRO\-CESS\_READ\-ER} for a non-shared reading end, and \textit{RE\-STRIC\-TED\_PRO\-CESS\_SHAR\-ED\_REA\-DER} for a shared reading end. Let us start with the \textit{RE\-STRIC\-TED\_PRO\-CESS\_WRIT\-ER} and the \textit{RE\-STRIC\-TED\_PRO\-CESS\_READ\-ER} from~\cite{ChalmersPedersen24} is shown below. We indicate where the claims and unclaims will be placed.
\begin{tabbing}
=\===\===\===\===\=================\==\==\==\= \kill
\>{\it RESTRICTED\_PROCESS\_WRITER}$(pid, chan)$ = \\
\>\>$schedule!pid\ \then$ \>\>\>\>{\tt --} schedule the process \\
\>\>$run.pid \then$ \>\>\>\>{\tt --} wait to be run \\ 
\>\>$running.pid.store!{\sf true} \then$\>\>\>\>{\tt --} {\bf this}.{\it running} = {\bf true};\\
\>\>{\it RESTRICTED\_PROCESS\_WRITER}$'(pid, chan)$\\
$ $\\
\>\>{\it RESTRICTED\_PROCESS\_WRITER}'$(pid, chan)$ =\\
\>\>$write.chan.pid?message \then$ \>\>\>\>{\tt --} take message from the env.\\
\>\>\underline{{\tt -\!\!\! -} place for future {\it CLAIM\_WRITE}}\\
\>\>$channel\_claim.chan!pid \then$ \>\>\>\>{\tt --} claim the channel\\
\>\>{\it WRITE}$(pid, chan, mesage);$ \>\>\>\>{\tt --} perform the write\\
\>\>$channel\_release.chan.pid \then$ \>\>\>\>{\tt --} release the channel\\
\>\>{\it YIELD}$(pid);$ \>\>\>\>{\tt --} $yield()$\\
\>\>\underline{{\tt -\!\!\! -} place for future {\it UNCLAIM\_WRITE}}\\
\>\>$ack.chan.pid \then$ \>\>\>\>{\tt --} acknowledge to the env.\\
\>\>{\it RESTRICTED\_PROCESS\_WRITER}'($pid, chan$) \\
$ $\\ 
\>{\it RESTRICTED\_PROCESS\_READER}($pid, chan$) =\\ 
\>\>$schedule.pid \then$ \>\>\>\>{\tt --} schedule the process\\
\>\>$run.pid \then$ \>\>\>\>{\tt --} wait to be run\\      
\>\>$running.pid.store!{\sf true}  \then$\>\>\>\>{\tt --} {\bf this}.{\it running} = {\bf true}\\
\>\>{\it RESTRICTED\_PROCESS\_READER}'($pid, chan$)\\
$ $\\
\>{\it RESTRICTED\_PROCESS\_READER}'($pid, chan$) =\\ 
\>\>$start\_read.chan.pid \then$ \>\>\>\>{\tt --} external event to start\\
\>\>\underline{{\tt -\!\!\! -} place for future {\it CLAIM\_READ}}\\
\>\>$channel\_claim.chan!pid \then$ \>\>\>\>{\tt --} claim the channel\\
\>\>$writer.chan.load?p \then$ \>\>\>\>{\tt --} get the {\it writer} field\\
\>\>(\\
\>\>\>${\sf if}\ (p == \mbox{\it NULL})\ {\sf then}$ \>\>\>{\tt --} {\bf if} ({\it writer} $==$ {\bf null}) {\bf then}    \\                     
\>\>\>\>$reader.chan.store.pid \then$ \>\>{\tt --}\>\>{\it reader} = {\bf this};\\
\>\>\>\>$ready.pid.store!false \then$ \>\>{\tt --}\>\>{\bf this}.{\it ready} = {\bf false};\\
\>\>\>\>$channel\_release.chan.pid \then$ \>\>{\tt --}\>\>release the channel\\
\>\>\>\>{\it YIELD}({\it pid}) \>\>{\tt --}\>\>$yield();$\\
\>\>\>{\sf else} \>\>\>{\tt --} {\bf else}\\
\>\>\>\>$channel\_release.chan.pid \then$  \>\>{\tt --}\>\>release the channel\\
\>\>\>\>{\sf SKIP}\\
\>\>);\\
\>\>$channel\_claim.chan!pid \then$ \>\>\>\>{\tt --} claim the channel\\
\>\>{\it READ}({\it pid, chan}); \>\>\>\>{\tt --} see {\it READ} below\\
\>\>$data.chan.load?message \then$ \>\>\>\>{\tt --} {\it message} = {\it data}\\
\>\>$channel\_release.chan.pid \then$ \>\>\>\>{\tt --} release the channel\\
\>\>(\\
\>\>\>$YIELD(pid);$\>\>\>{\tt --} $yield();$\\
\>\>\>\underline{{\tt -\!\!\! -} place for future {\it UNCLAIM\_READ}}\\

\>\>\>$read.chan.pid!message \then$ \>\>\>{\tt --} deliver msg to the env.\\
\>\>\>{\it RESTRICTED\_PROCESS\_READER}'({\it pid, chan}) \\
\>\>)
\end{tabbing}

To produce the shared versions of these procedures, all we have to do is insert {\it CLAIM\_\-WRITE/\-CLAIM\_\-READ} and {\it UN\-CLAIM\_\-WRITE/\-UN\-CLAIM\_\-READ} as indicated above.

\subsection{Building Shared Channels}

Now we can follow much the same procedure as we did when we created the specifications: for the many-to-one channel, interleave a number of writers and a single reader and run these in parallel with a single many-to-one channel. This can be done as follows:
\begin{tabbing}
=\===\===\===\===\=================\==\==\==\= \kill
\>{\it READER\_SHARED\_WRITER}({\it writers, R, C}) =\\
\>\>$(\;\;\mathclap{\underset{p \in \mbox{\it writers}}{|||}} \quad$ {\it RESTRICTED\_PROCESS\_SHARED\_WRITER}({\it p, C}))\\ 
\>\>$|||$\\
\>\>{\it RESTRICTED\_PROCESS\_READER}({\it R, C})\\
$ $\\
\>{\it RESTRICTED\_READER\_SHARED\_WRITER}({\it writers, R, C}) =\\
\>\>({\it READER\_SHARED\_WRITER}({\it writers, R, C}) \\ \>\>\hspace*{0.5cm}$\mathclap{\underset{\alpha{}\mbox{\it CHANNELS}}{||}} $\\ \>\>{\it MANY\_TO\_ONE\_CHANNEL}({\it C})) \ $\backslash \alpha{}${\it CHANNELS}
\end{tabbing}

If we were considering a classical unrestricted (unscheduled) system, {\it RESTRICTED\_READER\_SHARED\_WRITER} would be the implementation we would utilize in a verification scenario (though without the scheduling events), but since we are indeed taking into account the execution environment, we need to attach a scheduling system as well. We do this by simply running in parallel {\it RE\-STRICT\-ED\_READ\-ER\_SHAR\-ED\_WRIT\-ER} and {\it N\_SCHED\-ULER\_SYSTEM}.

\begin{figure}
\centerline{\includegraphics[height=8cm]{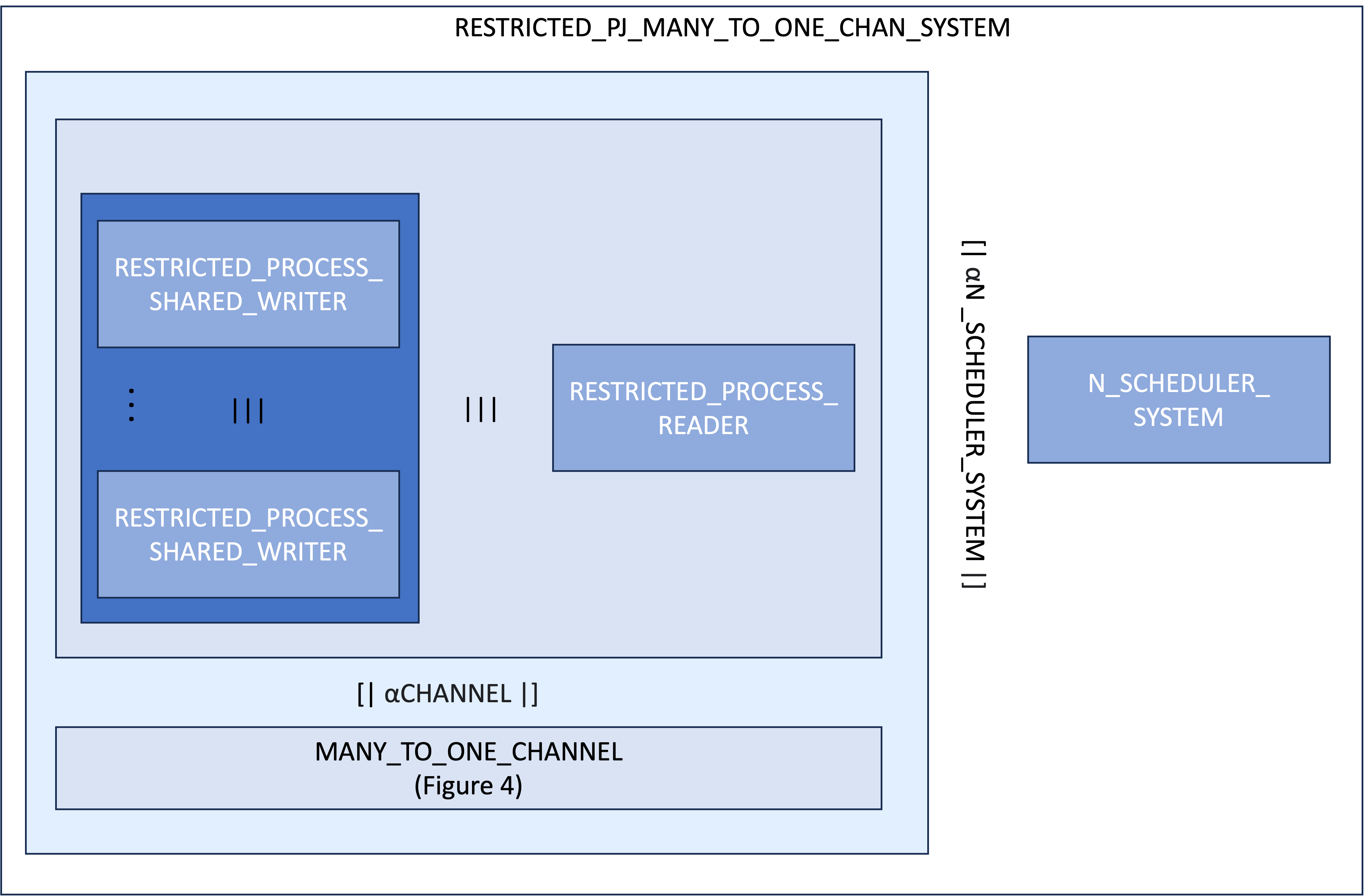}}
\caption{The {\it RESTRICTED\_PJ\_MANY\_TO\_ONE\_CHAN\_SYSTEM} process network.}
\label{fig:RestrictedPJMany2OneChanSystem}
\end{figure}

\begin{footnotesize}
\begin{tabbing}
=\===\===\===\===\=================\==\==\==\= \kill
{\it RESTRICTED\_PJ\_MANY\_TO\_ONE\_CHAN\_SYSTEM}({\it writers, R, C, N}) =\\ 
\>({\it RESTRICTED\_READER\_SHARED\_WRITER}({\it writers, R, C}) \\ \>\hspace*{2.3cm}$\mathclap{\underset{\alpha{}\mbox{\it N\_SCHEDULER\_SYSTEM}}{||}}$\\
\>{\it N\_SCHEDULER\_SYSTEM}($N$)) \ $\backslash \alpha{}${\it N\_SCHEDULER\_SYSTEM}
\end{tabbing}
\end{footnotesize}

\noindent
where $n$ is the number of schedulers. The equivalent one-to-many and many-to-many systems are built in a similar manner. The actual code can be found in the GitHub repository. Figure~\ref{fig:RestrictedPJMany2OneChanSystem} shows the process network of the {\it RESTRICTED\_PJ\_MANY\_TO\_ONE\_CHAN\_SYSTEM} process.

       % 7. CSP imp. of PJ shared chans
% !TeX root = ./Shared.tex
% Section 8
\section{A Generic Shared Channel Specification}
\label{sec:spec}

In CSP, a shared channel occurs when two or more processes interleave while they all are willing to communicate on a channel to another process. A simple example is:
\begin{tabbing}
=\===\===\===\===\================\==\==\= \kill
\>$P = a!x \then P$\\
\>$Q = a!y \then Q$\\
\>$R = a?z \then R$\\
\>{\it SYSTEM}$ = (P \ ||| \ Q) \quad \mathclap{\underset{ \{ a.v \ | \ v \in \mbox{\it Values} \} }{||}} \quad R$
\end{tabbing}
The channel $a$ synchronises $P$ and $R$ or $Q$ and $R$, but the process pair $P$ and $Q$ will not synchronise on channel $a$. In CSP, the behaviour is equivalent to a nondeterministic choice between $P$ and $Q$. For modelling, we must provide a specification that matches this behaviour.

In~\cite{PedersenChalmers} we presented a specification of a non-shared channel which originally was used by Welch and Martin in~\cite{WelchMartin00} to prove correctness of the channels from the JCSP Java package. They introduced a \textit{LEFT} and a \textit{RIGHT} process; the \textit{LEFT} process representing the writer and the \textit{RIGHT} process representing the reader. Our goal is to model the four possible channel interactions presented in Figure~\ref{fig:interactions}.

\begin{figure}
\centerline{\includegraphics[height=3cm]{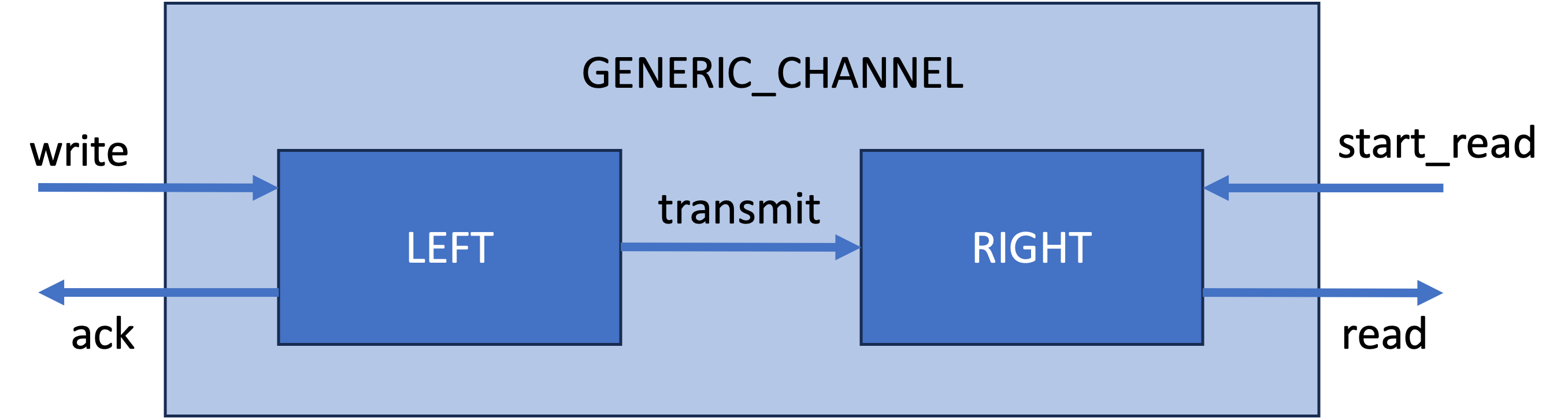}}
\caption{The composition of a {\it GENERIC\_CHANNEL with a {\it LEFT} writer and a {\it RIGHT} reader}.}\label{fig:LEFTRIGHT}
\end{figure}

\begin{tabbing}
=\===\===\===\===\================\==\==\= \kill
\>{\sf channel} \textit{transmit}: \textit{Channels}.\textit{Values}\\
$ $\\
\>{\it LEFT}$(pid, chan)$ = \\
\>\>$write.chan.pid?mess \then$ \>\>\>\>{\tt --} take a message\\
\>\>$transmit.chan!mess \then$ \>\>\>\>{\tt --} send it\\
\>\>$ack.chan.pid \then$       \>\>\>\>{\tt --} acknowledge the send\\
\>\>{\it LEFT}$(pid, chan)$ \>\>\>\> {\tt --} recurse\\
\\
\>{\it RIGHT}$(pid, chan)$ =\\
\>\>$start\_read.chan.pid \then\hspace{1cm}$ \>\>\>\>{\tt --} Start the read\\
\>\>$transmit.chan?mess \then$ \>\>\>\> {\tt --} received the message\\
\>\>$read.chan.pid!mess \then$ \>\>\>\> {\tt --} deliver the message\\
\>\>{\it RIGHT}$(pid, chan)$ \>\>\>\> {\tt --} recurse
\end{tabbing}
and a generic channel specification using these \textit{LEFT} and \textit{RIGHT} processes:
\begin{tabbing}
=\===\===\===\===\================\==\==\= \kill
\>{\it GENERIC\_CHANNEL}$(W, R, C) = $\\
\>\>$(${\it LEFT}$(W, C) \ _{\alpha{}\mbox{\it LEFT}} || _{\alpha{}\mbox{\it RIGHT}} \ ${\it RIGHT}$(R, C)) \ \backslash \ \{ |\ transmit\ | \}$
\end{tabbing}
Figure~\ref{fig:LEFTRIGHT} shows this process network. $\alpha$\textit{LEFT} and $\alpha$\textit{RIGHT} are the {\it alphabets} on which the two processes engage;  Their formal definitions are as follows:
\begin{tabbing}
=\===\===\===\===\================\==\==\= \kill
\>$\alpha{}LEFT(pid, chan) =$ \\
\>\>$\{write.chan.pid.v, transmit.chan.v, ack.chan.pid \mid v \leftarrow Values\}$\\
\>$\alpha{}RIGHT(pid, chan) =$\\
\>\>$\{start\_read.chan.pid, transmit.chan.v, read.chan.pid.v \mid \textit{v} \leftarrow Values\}$
\end{tabbing}

The \textit{transmit} event is hidden\footnote{We hide the complete set of events for {\it transmit}, denoted by the set $\{ | transmit | \}$.} and the \textit{LEFT} process engages with the environment on \textit{write} and \textit{ack} (the former representing the start of a {\it write} call on a channel and the latter representing the completion, and the \textit{RIGHT} process on the events \textit{start\_read} and \textit{read} (the former representing that start of a {\it read} call on a channel and the latter representing the completion and return of the read value). 

Recall, ProcessJ has the following three different types of shared channels:
\begin{itemize}
\item A many-to-one channel (multiple writers to one reader).
\item A one-to-many channel (one writer to multiple readers).
\item A many-to-many channel (multiple writers to multiple readers).
\end{itemize}

\begin{figure}
\centerline{\includegraphics[height=8cm]{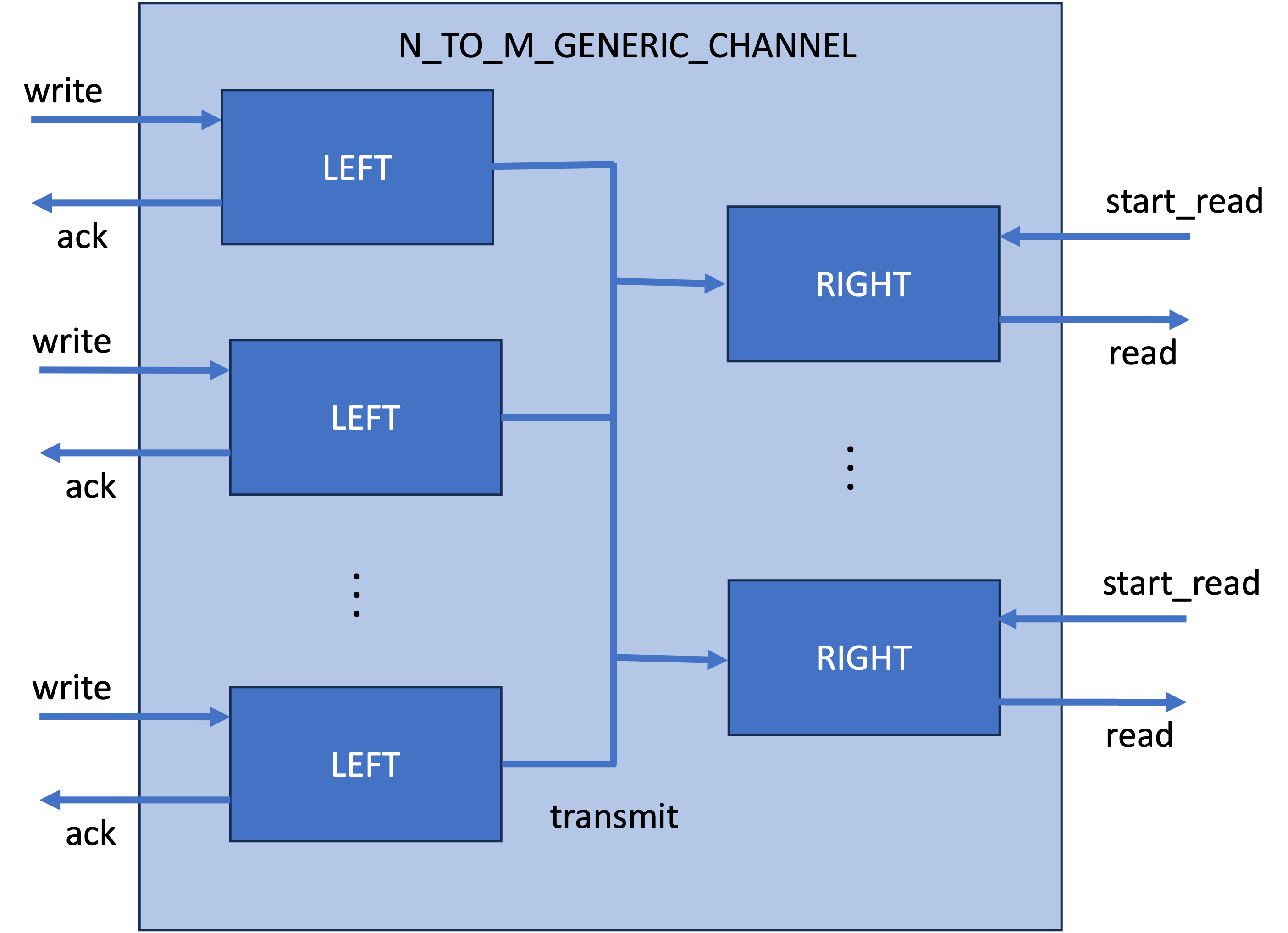}}
\caption{{\it N\_TO\_M\_GENERIC\_CHANNEL network.}}
\label{fig:N2MGenericChannel}
\end{figure}

We can specify all of these using the {\it LEFT} and the {\it RIGHT} processes described above. Let us start with the more general of the three, namely, the many-to-many (multiple writers and multiple readers); we call this one {\it N\_TO\_M\_GENERIC\_CHANNEL} and it looks like this:

\begin{tabbing}
=\===\===\===\===\================\==\==\= \kill
\>{\it N\_TO\_M\_GENERIC\_CHANNEL}$(writers, readers, C) = $\\
\>\>$((\underset{p \in writers}{|||} LEFT(p, C))$\\
\>\>$\underset{ \{ | transmit | \} }{||}$\\
\>\>$(\underset{p \in readers}{|||} RIGHT(p, C))) \ \backslash \ \{ | transmit | \}$
\end{tabbing}

\noindent
where {\it writers} is a list of $n$ process identifiers for the writing processes (i.e., the {\it LEFT} processes), and {\it readers} is a list of $m$ process identifiers for the reading processes (i.e., the {\it RIGHT} processes). The $n$ writers and $m$ readers all use the same {\it transmit} channel (the actual communication channel between left and right sides), and since all the readers are interleaved and all the writers are interleaved, only one writer and one reader will be paired up and communicate using the {\it transmit} channel. Thus, a single communication happens across the {\it transmit} channel, which is finally hidden from the environment. Figure~\ref{fig:N2MGenericChannel} illustrates the {\it N\_TO\_M\_GENERIC\_CHANNEL} generic channel network. Note, each {\it write/ack} and each {\it start\_read/read} pair of channels belong to different processes. Channel communication between a reader-writer pair still behaves as illustrated in Figure~\ref{fig:interactions}.

The one-to-many and the many-to-one channels can easily be specified using the {\it N\_TO\_M\_GENERIC\_CHANNEL}. Simply pass a list with a single writer to get a one-to-many channel, and to get a many-to-one channel, pass a list with a single reader.

In the next section, we follow the procedure of Welch and Martin~\cite{WelchMartin00} by proving that our shared channel specification indeed behaves like a shared channel in CSP. Having accomplished this, we know that it is safe to use our specification of a shared channel in the subsequent CSP code.             % 8. A Generic Channel Spec 
% !TeX root = ./Shared.tex
% Section 9
\section{Results}
\label{sec:results}

With both the CSP for the {\it specification} of a shared channel (Section~\ref{sec:spec}) and the {\it implementation} (Section~\ref{sec:imp}), we can now perform the actual verification using FDR.

We start by checking for divergence and deadlock freedom in both the specification and the implementation (with 1, 2, 3, and 4 schedulers); indeed both the specification and the implementation are divergence and deadlock free. Checking both the specification and the implementation for determinism fails---as it should. This means that we can limit our refinement checks to the {\it stable failures} model. 

\newcolumntype{x}[1]{>{\centering\arraybackslash\hspace{0pt}}p{#1}}
\begin{table}
\caption{Results for the one-to-many channel. (F = Failures, T = Traces)}
\label{tab:results}
\begin{center}
\begin{adjustbox}{angle=0}
\begin{tabular}{|c|c|x{1cm}|x{1cm}|x{1cm}|x{1cm}|}\hline
\textit{Spec} $\refinedby$ \textit{Impl} & \# Active & \multicolumn{4}{c|}{Number of schedulers}  \\ 
                                         & Procs. &\textbf{1} & \textbf{2} & \textbf{3} & \textbf{4}\\ \hline
$writers-readers$&\multicolumn{5}{c|}{One-to-many}\\ \hline
1-2    & 3 & T & T & F & F \\ 
1-3    & 4 & T & T & T & F \\ \hline
$writers-readers$& \multicolumn{5}{c|}{Many-to-one}\\ \hline
2-1    & 3 & T & T & F & F \\
3-1    & 4 & T & T & T & F\\ \hline
$writers-readers$&\multicolumn{5}{c|}{Many-to-many}\\ \hline
2-2    & 4 & T & T & T & F \\\hline\hline
\textit{Spec} $\refinedby$ \textit{Impl} & \# Active & \multicolumn{4}{c|}{Number of schedulers}  \\ 
                                         & Procs. &\textbf{1} & \textbf{2} & \textbf{3} & \textbf{4}\\ \hline
$writers-readers$&\multicolumn{5}{c|}{One-to-many}\\ \hline  
1-2    & 3 & \xmark &\xmark & F & F \\ 
1-3    & 4 &\xmark&\xmark&\xmark&F\\ \hline
$writers-readers$& \multicolumn{5}{c|}{Many-to-one}\\ \hline
2-1    & 3 & \xmark & \xmark & F & F \\
3-1    & 4 & \xmark & \xmark & \xmark & F\\\hline
$writers-readers$& \multicolumn{5}{c|}{Many-to-many}\\ \hline
2-2    & 4 & \xmark & \xmark & \xmark & F\\ \hline
\end{tabular}
\end{adjustbox}
\end{center}
\end{table}

Table~\ref{tab:results} shows the results of verification of the three different ProcessJ shared channel types.

Note, for the one-to-many we have one writer, and two or three readers---a total of three or four processes. Similarly, the many-to-one has two or three writers and a single reader; again, three or four processes. 

We see that when the number of schedulers is three for these two channel types, the results are positive for one writer and two readers as well as for two writers and one reader in the stable failures model in both directions. When it comes to three writers and a reader or one writer and three readers, the required number of schedulers to achieve refinement in the failures model is four.

For the many-to-many channel with two writers and two readers---a total of four processes, refinement in the failures model is not achieved until the number of schedulers equals at least four. In general, with $n$ writers and $m$ readers, a total of $n+m$ schedulers will be needed to obtain refinement in both directions in the stable failures model.

With the results from this paper, we now have verified the scheduled implementations of the processJ runtime for the following entities:
\begin{itemize}
\item One-to-one shared channels~\cite{PedersenChalmers}.
\item Choice (alts) for up to 3 writers~\cite{ChalmersPedersen24}.
\item One-to-many, many-to-one, and many-to-many channels [this paper].
\end{itemize}
In Section~\ref{sec:discussion}, we will look closer at the results that did not pass the refinement check, and explain why.         % 9. Results
%!TeX root = shared.tex
% Section 10
\section{Formal proof of shared channel spec behaving like a shared CSP channel}
\label{sec:proof}

We begin as Welch and Martin by considering ALT-free CSP programs. We will start with one that uses a many-to-one channel with no alternation. We define a $SCSP$ network as a special kind of CSP network $P_1 \; || \; P_2 \; || \; \dots \; || \; P_n$ where $P_i$ is defined as a $SCSPPROC$:
\begin{align*}
    SCSPPROC    &= SKIP \\
                &\ \ | \;\ a!x \to SCSPPROC \\
                &\ \ | \;\ a?x \to SCSPPROC \\
                &\ \ | \;\ SCSPPROC \; \Box \; SCSPPROC \\
                &\ \ | \;\ SCSPPROC \; \sqcap \; SCSPPROC \\
                &\ \ | \;\ SCSPPROC \; ||| \; SCSPPROC
\end{align*}

We have added $SCSPPROC \; ||| \; SCSPPROC$ to Welch and Martin's original proof. For a many-to-one channel with two writers, we have the following specification:

\begin{tabbing}
=\===\===\===\===\================\==\==\= \kill
\>{\it MANY\_TO\_ONE\_CHANNEL}$(a) =$\\
\>\>((({\it LEFT}($a$) ||| {\it LEFT}($a$)) || {\it RIGHT}($a$)))  $\backslash  \{ a \}$
\end{tabbing}
where
\begin{tabbing}
=\===\===\===\===\================\==\==\= \kill
\>{\it LEFT}$(a) = write.a?x \then a!x \then ack.a \then$ {\it LEFT}($a$) \\
\>{\it RIGHT}$(a) = start\_read.a \then a?z \then read.a!z \then$ {\it RIGHT}($a$)
\end{tabbing}
We only need to prove the case for two {\it LEFT} processes as the proof simply expands for more {\it LEFT} processes as we will show. We now need to demonstrate that this channel does indeed behave as a shared CSP channel as presented in Section~\ref{sec:spec}. We will examine an SCSCP network $V = P_1 \; || \; \dots \; || \; P_n$ and algebraically transform it so the CSP channel is replaced with our {\it MANY\_TO\_ONE\_CHANNEL}.

Our transformation replaces each channel communication with the equivalent {\it MANY\_TO\_ONE\_CHANNEL}. We undertake the following steps:
\begin{enumerate}
  \item We replace a CSP channel $a$ in $V$ with {\it MANY\_\-TO\_\-ONE\_\-CHAN\-NEL}.
  \item We transform the process $P_i$ to $P'_i$ by replacing any $a!x \then$ {\it Process} by $write.a!x \then ack.a \then$ {\it Process}.
  \item We transform the process $P_j$ to $P'_j$ by replacing any $a?x \then$ {\it Process} by $start\_read.a \then read.a?x \then$ {\it Process}.
\end{enumerate}

We will show that the external behaviour of subnetwork $P_i || P_j$ is unchanged by this transformation. We need to prove that:
\begin{tabbing}
=\===\===\===\===\================\==\==\= \kill
\>$(P_i \; || \; P_j) \; \backslash \; \{ a \} =$\\
\>\>$(P'_i \; || \; MANY\_TO\_ONE\_CHANNEL(a) \; || \; P'_j) \;$\\
\>\>$\backslash \; \{ write.a, ack.a, start\_read.a, read.a \}$
\end{tabbing}
Starting with the RHS, we replace {\it MANY\_TO\_ONE\_CHANNEL} with two {\it LEFT} and one {\it RIGHT} processes, hiding the internal $a$.
\begin{tabbing}
=\===\===\===\===\================\==\==\= \kill
\>$(P'_i \; || \; \mbox{\it MANY\_TO\_ONE\_CHANNEL}(a) \; || \; P'_j) $\\
\>\>\>$\backslash \; \{ write.a, ack.a, start\_read.a, read.a \} =$ \\
\>$(P'_i \; || \; (((\mbox{\it LEFT}(a) \; ||| \; \mbox{\it LEFT}(a)) \; || \; \mbox{\it RIGHT}(a)) \; \backslash \; \{ a \}) \; || \; P'_j)$\\
\>\>\>$\backslash \; \{ write.a, ack.a, start\_read.a, read.a \}$
\end{tabbing}

As $P'\_i$ and $P'\_j$ have had all occurrences of $a$ we can move the internal hiding set to the outside of the definition.
\begin{tabbing}
=\===\===\===\===\================\==\==\= \kill
\>$(P'_i \; || \; (((\mbox{\it LEFT}(a) \; ||| \; \mbox{\it LEFT}(a)) \; || \; \mbox{\it RIGHT}(a)) \backslash \; \{ a \}) \; || \; P'_j)$\\
\>\>$\backslash \; \{ write.a, ack.a, start\_read.a, read.a \} =$ \\
\>$(P'_i \; || \; ((\mbox{\it LEFT}(a) \; ||| \; \mbox{\it LEFT}(a)) \; || \; \mbox{\it RIGHT}(a)) \; || \; P'_j)$\\ 
\>\>$\backslash \; \{ a, write.a, ack.a, start\_read.a, read.a \}$
\end{tabbing}
Also, as the $write$ and $ack$ pair and the $start\_read$ and $read$ pair only exist in $P'_i$ and $P'_j$ respectively, we can rewrite the hiding as follows:
\begin{tabbing}
=\===\===\===\===\================\==\==\= \kill
\>$(P'_i \; || \; ((\mbox{\it LEFT}(a) \; ||| \; \mbox{\it LEFT}(a)) \; || \; \mbox{\it RIGHT}(a)) \; || \; P'_j)$\\
\>\>$\backslash \; \{ a, write.a, ack.a, start\_read.a, read.a \} = $\\
\>$(((P'_i \; || \; (\mbox{\it LEFT}(a) \; ||| \; \mbox{\it LEFT}(a))) \; \backslash \; \{ write.a, ack.a \}) \; ||
  \; ((\mbox{\it RIGHT}(a) \; || \; P'_j)$\\
\>\>$\backslash \; \{ start\_read.a, read.a \})) \; \backslash \; \{ a \}$
\end{tabbing}
We therefore need to prove the following:
\begin{align}
  (P'_i \; || \; (\mbox{\it LEFT}(a) \; ||| \; \mbox{\it LEFT}(a))) \; \backslash \; \{ write.a, ack.a \} &= P_i \; \text{and}\\
  (\mbox{\it RIGH}T(a) \; || \; P'_j) \; \backslash \; \{ start\_read.a, read.a \} &= P_j
\end{align}
$P_i$ is the interleaving of two writing processes on the channel $a$, meaning $P_i = (\mu q.a!x \to q) \; ||| \; (\mu r.a!x \to r)$. $P_j$ is a single reading process, meaning $P_j = a?x \to P_j$.

For equation 1, we will distinguish between the two {\it LEFT}($a$) processes as {\it LEFT} and {\it LEFT'}:
\begin{tabbing}
=\===\===\===\===\================\==\==\= \kill
\>{\it LEFT}$(a) = write.a!x \then ack.x \then$ {\it LEFT}($a$) \\
\>{\it LEFT'}$(a) = write.a!y \then ack.y \then$ {\it LEFT'}($a$)
\end{tabbing}
With our restricted SCSP syntax, we know equation 1 is true as:

\begin{enumerate}
  \item When transforming $P_i$ to $P'_i$ we replaced $a!x \then$ {\it PROCESS} with $write.a!x \then ack.a \then$ {\it PROCESS}.
  \item The parallel combination of {\it LEFT}$(a) \; ||| \; \mbox{\it LEFT'}(a)$ and $P'_i$ produces the process $(\mu q.write.a!x \to a!x \to ack.a \to q) \; ||| \; (\mu r.write.a!y \to a!y \to ack.a \to r)$.
  \item Hiding $\{ write.a, ack.a \}$ of the process $(\mu q.write.a!x \to a!x \to ack.a \to q) \; ||| \; (\mu r.write.a!y \to a!y \to ack.a \to r)$ gives us $(\mu q.a!x \to q) \; ||| \; (\mu r.a!y \to r)$ which is the same definition of $P_i$.
\end{enumerate}

For equation 2, we follow the same steps thus concluding that the transformation to replace a CSP channel with a {\it MANY\_\-TO\_\-ONE\_\-CHAN\-NEL}(a) did not affect the external behaviour of the network. We end up with our original example:
\begin{align*}
  ((\mu q.a!x \to q) \; ||| \; (\mu r.a!y \to r)) \; || \; (\mu p.a?z \to p)
\end{align*}
As can be seen, it would be easy to increase the number of {\it LEFT} processes to any number as the transformation provides demonstrates these processes become an interleaving of writing processes.

A similar process can be undertaken to demonstrate the equivalence of a {\it ONE\_TO\_MANY\_CHANNEL} by introducing two {\it RIGHT} processes. {\it MANY\_TO\_MANY\_CHANNEL} also follows a similar process but with two {\it LEFT} and {\it RIGHT} processes. Note, in the CSP model the exclusive access to the {\it transmit} channel is ensured by reading and writing processes being {\it interleaved} and the {\it transmit} channel access being nondeterministic.     % 10. Formal Proof
%!TeX root = ./Shared.tex
% Section 11
\section{Discussion}
\label{sec:discussion}

In this paper we have proved that the modelled implementation of shared channels in ProcessJ will behave correctly if implemented according to this model:
\begin{itemize}
\item We get failures refinement (and there is never any divergence) if the number of schedulers is equal to or greater than the number of processes.
\item If less, we only get trace-refinement because some traces are no longer possible. This does not mean that the system will deadlock or behave inappropriately, it just means that some traces are no longer realisable.
\end{itemize}  

There are two questions that need answering, namely, why do we only get trace refinement in the $Spec \refinedby Impl$ case and why does refinement checking fail for some cases in the $Impl \refinedby Spec$ scenario.

Let us tackle the first question: the run queue that ready-to-run processes go into naturally imposes an order on processes. Remember, processes in the run queue are ready to run; all they need is a scheduler to run them. If we consider the simple case with just two processes, say, $P_A$ and $P_B$ in the run queue like this: $[P_A, P_B]$, and two schedulers, then it should be reasonably clear that these schedulers each can get a process to run. All possible interleavings of events from running $P_A$ and $P_B$ concurrently are possible. However, if we only have a single scheduler available, only $P_A$ will be executed and only events from this execution are observable. However, once $P_A$ yields, the scheduler will run $P_B$ and we can observe all its events. In reality, in the latter case the fact that both $P_A$ and $P_B$ could be run concurrently is of no consequences and they end up being run {\bf in sequence} as there is only one scheduler to handle both processes, but only one at a time. $P_A$ and $P_B$ both have acceptance sets; that is, the set of events they are willing to engage on after a certain trace. If there are schedulers enough to run all ready processes, then the system's overall acceptance set is the union of the acceptance sets of every process in the system. However, if, as above, there are not enough schedulers, then any process that does not have a scheduler will not have its acceptance set included in the overall system's acceptance set. In a non-scheduled CSP system this is never a problem as events can happen when they are ready---they do not require any scheduler or runner code to execute them. This explains why failures refinement can never happen when there are not enough schedulers to run every ready process.
At this point it is worth noting that even though we do get traces refinement, the serialisation that happens when not enough schedulers are available also reduces the number of possible traces. That is, a scheduled system with less schedulers than processes may result in some traces never being possible. 

Now, once we realise that it becomes clear that when checking refinement in the other direction ($Impl \refinedby Spec$) fails---even in the traces model---since there are now traces that the $Spec$ can perform that the $Impl$ never could, simply because of a lack of schedulers to maximise concurrency in the execution. CSP does not have any notion of ``the order of processes being executed''; events happen instantaneously. However, that is not how things work in the real world. Code needs to be run and synchronising on events takes time.
 
%the thing about queues: if execution order is serialized by for example a queue, a certain order {\bf will} be imposed on processes next to each other in the queue if there is not enough schedulers to run all those processes at the same time. A safe upper bound for the number of schedulers required to obtain failures refinement in both directions is the number of processes in the system. However, it is possible that fewer schedulers will suffice: the lower bound for failures refinement in both directions is the largest number of processes ever ready to run at the same time; in other words, when a process is ready to run, it will be run immediately by a scheduler.       % 11. Discussion
% !TeX root = ./Shared.tex
% Section 12
\section{Future Work}
\label{sec:futureWork}

Surely speed and efficiency in a concurrent system is a design goal, but so is correctness. After all, a system that is incorrect is of no use to any one. Looking at the proposed implementation of shared channels presented in this paper, we note the frequent use of locks/monitors. We have locks on:
\begin{itemize}
\item Entire processes. This is currently necessary to correctly handle setting the {\it ready} and {\it running} flags and to correctly place processes in the run queue.
\item Entire channels. This is necessary in order to avoid race conditions when communicating on the channel.
\end{itemize}

Locking data reduces concurrency, which will impact the speed with which the code runs. Therefore, we want to consider removing the use of locks and replacing them with atomic data structures. To investigate this potential improvement, we need to develop CSP models for atomic variables and a wait-free queue:
\begin{itemize}
\item  Atomic variables for the fields which  support compare-and-swap. That is, a compare-and-swap operation allows for comparing the variable (v) to a value (m) and, if the comparison succeeds, set the v to a new value (n).
\item A wait-free-queue to replace the current queues in the runtime. A wait-free queue allows multiple processes to access it safely at the same time.
\end{itemize}

Given that the ProcessJ implementation of channels does not meet the specification without enough scheduling resources, we do not have a general, lightweight construct to build CSP models of ProcessJ systems. To model systems, we will require a simplified specification of how ProcessJ performs channel communication. This simplification can be tested against our ProcessJ implementation models to ensure they behave as desired. Such simplified models can then be used to build concurrent systems to analyse how ProcessJ will behave under different scheduling conditions for different scheduling problems. Of further interest is investigating the number of schedulers needed to ensure failures refinement in both directions. We have shown that if every process gets its own scheduler, then failures refinement holds in both directions, but is it possible that we  can do with a number less that the number of processes in the system. This is an interesting topic to investigate and naturally, the answer will depend on the program being run.

          % 12. Future Work

%\begin{biography}{\includegraphics[width=76pt,height=76pt,draft]{empty}}{
%{\textbf{Jan B\ae{}kgaard Pedersen.} Please check with the journal's %author guidelines whether
%author biographies are required. They are usually only included for
%review-type articles, and typically require photos and brief
%biographies for each author.}}
%\end{biography}

\bibliographystyle{elsarticle-num-names}
\bibliography{biblio.bib}

\end{document}